\documentclass[11pt]{article}
\pdfoutput=1  
\usepackage{graphicx,color}
\usepackage{tikz}
\usetikzlibrary{decorations.pathmorphing}	
\usetikzlibrary{decorations.markings}
\tikzset{
    graviton/.style={decorate, decoration={snake,amplitude=2.5pt}, draw},
    healthy/.style={solid,draw=black},
    ghost/.style={dashed,draw=black},
}

\usepackage{appendix}
\usepackage{latexsym,amsmath,amssymb,graphicx,booktabs}
\usepackage{epsfig,latexsym,cite}
\usepackage{hyperref}
\numberwithin{equation}{section}

\definecolor{MyBlue}{rgb}{0.15,0.15,0.70}

\hypersetup{
colorlinks=true,
citecolor=MyBlue,
linkcolor=MyBlue,
urlcolor=MyBlue
}

\input epsf
\usepackage{amssymb}
\usepackage{amsmath}
\usepackage{amsfonts}
\usepackage{upgreek}
\usepackage{latexsym}

\newcommand{\nn}{\nonumber}

\newcommand{\iBox}{\Box^{-1}}

\renewcommand\({\left(}
\renewcommand\){\right)}
\renewcommand\[{\left[}
\renewcommand\]{\right]}

\newcommand\n{{\mbox {\boldmath $\nabla$}}}
\newcommand{\ra}{\rightarrow}

\def\lsim{\raise 0.4ex\hbox{$<$}\kern -0.8em\lower 0.62
ex\hbox{$\sim$}}

\def\gsim{\raise 0.4ex\hbox{$>$}\kern -0.7em\lower 0.62
ex\hbox{$\sim$}}

\def\lbar{{\hbox{$\lambda$}\kern -0.7em\raise 0.6ex
\hbox{$-$}}}

\newcommand\eq[1]{eq.~(\ref{#1})}
\newcommand\eqs[2]{eqs.~(\ref{#1}) and (\ref{#2})}
\newcommand\Eq[1]{Equation~(\ref{#1})}

\newcommand\eqst[2]{eqs.~(\ref{#1})--(\ref{#2})}

\newcommand\pa{\partial}
\newcommand\p{\partial}

\newcommand\ee{\end{equation}}
\newcommand\be{\begin{equation}}
\def\bea{\begin{array}}
\def\eea{\end{array}}\def\ea{\end{array}}
\newcommand\ees{\end{eqnarray}}
\newcommand\bees{\begin{eqnarray}}
\def\nn{\nonumber}

\def\a{\alpha}
\def\b{\beta}

\def\g{\gamma}

\def\d{\delta}

\def\eps{\epsilon}

\def\dslash{\hspace{-1mm}\not{\hbox{\kern-2pt $\partial$}}}
\def\Dslash{\not{\hbox{\kern-4pt $D$}}}
\def\pslash{\not{\hbox{\kern-2.1pt $p$}}}
\def\kslash{\not{\hbox{\kern-2.3pt $k$}}}
\def\qslash{\not{\hbox{\kern-2.3pt $q$}}}

\newcommand{\vk}{{\bf k}}
\newcommand{\vx}{{\bf x}}

\def\p1{{\bf p}_1}
\def\p2{{\bf p}_2}
\def\k1{{\bf k}_1}
\def\k2{{\bf k}_2}

\newcommand{\emn}{\eta_{\mu\nu}}

\newcommand{\eMN}{\eta^{\mu\nu}}

\newcommand{\gmn}{g_{\mu\nu}}

\newcommand{\gMN}{g^{\mu\nu}}

\newcommand{\gbmn}{\bar{g}_{\mu\nu}}

\newcommand{\hmn}{h_{\mu\nu}}

\newcommand{\xim}{\xi_{\mu}}
\newcommand{\xin}{\xi_{\nu}}

\newcommand{\pam}{\pa_{\mu}}

\newcommand{\pan}{\pa_{\nu}}

\newcommand{\paM}{\pa^{\mu}}

\newcommand{\Rmn}{R_{\mu\nu}}
\newcommand{\Gmn}{G_{\mu\nu}}
\newcommand{\RMN}{R^{\mu\nu}}

\newcommand{\Rmnrs}{R_{\mu\nu\rho\sigma}}
\newcommand{\RMNRS}{R^{\mu\nu\rho\sigma}}

\newcommand{\Tmn}{T_{\mu\nu}}

\newcommand{\TMN}{T^{\mu\nu}}

\newcommand{\dddM}{\kern 0.2em \raise 1.9ex\hbox{$...$}\kern -1.0em \hbox{$M$}}
\newcommand{\dddQ}{\kern 0.2em \raise 1.9ex\hbox{$...$}\kern -1.0em \hbox{$Q$}}
\newcommand{\dddI}{\kern 0.2em \raise 1.9ex\hbox{$...$}\kern -1.0em\hbox{$I$}}
\newcommand{\dddJ}{\kern 0.2em \raise 1.9ex\hbox{$...$}\kern-1.0em
\hbox{$J$}}
\newcommand{\dddcalJ}{\kern 0.2em \raise 1.9ex\hbox{$...$}\kern-1.0em
\hbox{${\cal J}$}}

\newcommand{\dddO}{\kern 0.2em \raise 1.9ex\hbox{$...$}\kern -1.0em
\hbox{${\cal O}$}}
\def\dddz{\raise 1.5ex\hbox{$...$}\kern -0.8em \hbox{$z$}}
\def\dddd{\raise 1.8ex\hbox{$...$}\kern -0.8em \hbox{$d$}}
\def\dddbd{\raise 1.8ex\hbox{$...$}\kern -0.8em \hbox{${\bf d}$}}
\def\ddbd{\raise 1.8ex\hbox{$..$}\kern -0.8em \hbox{${\bf d}$}}
\def\dddx{\raise 1.6ex\hbox{$...$}\kern -0.8em \hbox{$x$}}

\newcommand{\mpl}{M_{\rm Pl}}
\newcommand{\mplr}{m_{\rm Pl}}

\newcommand{\ola}{\Omega_{\Lambda}}

\begin{document}

\begin{titlepage}

\vspace*{2cm}

\centerline{\Large \bf Dark energy and dimensional transmutation in $R^2$ gravity}

%

\vskip 0.4cm
\vskip 0.7cm
\centerline{\large Michele Maggiore}
\vspace{3mm}
\centerline{\em D\'epartement de Physique Th\'eorique and Center for Astroparticle Physics,}  
\centerline{\em Universit\'e de Gen\`eve, 24 quai Ansermet, CH--1211 Gen\`eve 4, Switzerland}

\vskip 1.9cm

\begin{abstract}

Recent work has shown that non-local modifications of gravity involving terms such as $m^2R\Box^{-2}R$ (and no cosmological constant) provide a phenomenologically viable alternative to $\Lambda$CDM. We first discuss the possibility that such non-local terms emerge in the far infrared from the running of a coupling constant associated to the $R^2$ term in higher-derivative gravity, which, depending on the UV completion of the theory, can be  asymptotically free in the ultraviolet and strongly coupled in the infrared. In this scenario  the mass scale $m$  of the non-local model emerges from dimensional transmutation, similarly to  $\Lambda_{\rm QCD}$ for strong interactions,   leading to a technically natural value and to a novel understanding of the scale associated to dark energy.  Motivated by these findings, we  then explore the possibility of generating strong infrared effects in Einstein gravity, with no $R^2$ terms, as a consequence of the higher-derivative term generated by the conformal anomaly.

\end{abstract}

\end{titlepage}

\newpage

\section{Introduction}

At the fundamental level, quantum field theory is local. Still,  in many situations non-locality emerges as a derived phenomenon. This can happen already  at a purely classical level, when one
integrates out some fast degree of freedom to obtain an effective theory for the slow degrees of freedom, or at the quantum level, when one considers the effective action that takes into account loop corrections involving light or massless particles. In recent years there  has been a significant activity on effective non-local modifications of gravity, largely motivated by the aim of understanding the origin of dark energy.
In particular, in \cite{Maggiore:2013mea} (elaborating on previous works related to
the degravitation idea~\cite{ArkaniHamed:2002fu,Dvali:2006su,Dvali:2007kt}, as well as on attempts at writing massive gravity in non-local form~\cite{Porrati:2002cp,Jaccard:2013gla})
we proposed a phenomenological  modification of gravity, based on
the non-local equation of motion
\be\label{RT}
\Gmn -(1/3)m^2\(\gmn\iBox R\)^{\rm T}=8\pi G\,\Tmn\, .
\ee
The superscript T denotes the operation of taking the transverse part of a tensor (which is itself a nonlocal operation), $\Box$ is the covariant d'Alembertian computed with the curved-space metric $\gmn$, and its inverse
$\iBox$ is defined using the retarded Green's function, to ensure causality. The factor $1/3$ is a convenient normalization of the parameter $m^2$ in $d=3$ spatial dimensions. The extraction of the transverse part ensures that energy-momentum conservation is  automatically satisfied. A closed form for the action corresponding to \eq{RT} is not known. This model is however  closely related to another non-local model, proposed in \cite{Maggiore:2014sia}, and defined by the action
\be\label{RR}
S_{\rm NL}=\frac{\mplr^2}{2}\int d^{4}x \sqrt{-g}\, 
\[R-\frac{1}{6} m^2R\frac{1}{\Box^2} R\]\, .
\ee
These two models are related by the fact that,  linearizing over flat space the equations of motion derived from the action (\ref{RR}), one finds the same equations of motion as those obtained by linearizing \eq{RT} over flat space. However, at the full non-linear level, or linearizing over a background different from Minkowski, the two theories are different. 

An intriguing aspect of the  models (\ref{RT}) and (\ref{RR}) is that, in the comparison with cosmological observations,  they perform remarkably well. They both have the same number of parameters as $\Lambda$CDM, with the mass $m$ replacing the cosmological constant. They dynamically generate  a dark energy and have a realistic background FRW evolution \cite{Maggiore:2013mea,Maggiore:2014sia,Foffa:2013vma}. Their cosmological perturbations are well-behaved (a step that already ruled out several modified gravity models)
and their quantitative effects are consistent with CMB, supernovae, BAO and structure formation data~\cite{Nesseris:2014mea,Dirian:2014ara,Barreira:2014kra}.\footnote{This should be contrasted with the non-local model proposed in \cite{Deser:2007jk,Deser:2013uya,Woodard:2014iga}. This model does not involve a mass scale, and is rather 
constructed adding to the Einstein-Hilbert action a term of the form $Rf(\iBox R)$. The function $f(\iBox R)$ is  tuned so that, at the level of background evolution, this model closely mimics $\Lambda$CDM. One can then study its
cosmological perturbations, and  it has been found in 
\cite{Dodelson:2013sma} that the  model is ruled out  by the comparison with structure formation.
Non-local long-distance modifications of GR have also been suggested in \cite{Barvinsky:2003kg,Barvinsky:2011hd,Barvinsky:2011rk,Wetterich:1997bz}.}
This allowed  a more detailed comparison with $\Lambda$CDM. Implementing the cosmological perturbations of the non-local models into a Boltzmann code and performing parameter estimation and a global fit to CMB, supernovae and BAO data, one finds that these models perform as well as $\Lambda$CDM, with  comparable values of the $\chi^2$   \cite{Dirian:2014bma} (in fact for model (\ref{RT}) the $\chi^2$ is even better than for $\Lambda$CDM, although not at a statistically significant level). Furthermore, parameter estimation provides a value of the Hubble constant $H_0$ slightly higher than in $\Lambda$CDM, in better agreement with that obtained from the most recent local measurements \cite{Rigault:2014kaa}. 
To the best of our knowledge, among the large
variety of existing  alternatives to the standard $\Lambda$CDM  paradigm,
the models defined by  eqs.~(\ref{RT}) or (\ref{RR}) are the only  ones
that are competitive with $\Lambda$CDM from the point of view of fitting current observations (at a level of accuracy which tests not only the background evolution but also the cosmological perturbations of the model), without being just an extension of $\Lambda$CDM with extra free parameters (see also \cite{Ade:2015rim}). Further conceptual and phenomenological aspects of these models have been discussed in \cite{Foffa:2013sma,Kehagias:2014sda,Cusin:2014zoa,Dirian:2014xoa,Modesto:2013jea,Conroy:2014eja,Barreira:2015fpa,Barreira:2015vra}.

These non-local models have been proposed on a purely phenomenological ground, and it is of course important to eventually understand how they could emerge from a fundamental theory. The purpose of this paper is to investigate the possibility that such terms emerge from   infrared (IR) effects in  gravity. We will first discuss  the idea of generating the dark-energy scale from the  running of asymptotically-free coupling constants in $R^2$ extensions of Einstein gravity. We will then explore the possibility of generating similar infrared effects in Einstein gravity, with no $R^2$ terms, just as a consequence of the higher-derivative terms generated by the conformal anomaly.

The plan of the paper is as follows. In Section~\ref{sect:one} we recall known results on the running of the couplings associated to the ${\cal O}(R^2)$ terms. In sect.~\ref{sect:IR} we discuss the possible IR dynamics that could be generated by the running of these coupling constants, and the  connection with the non-local models (\ref{RT}) and (\ref{RR}). Just as in QCD, the running of a coupling that becomes strong in the IR generates,  by dimensional transmutation, a mass scale $\Lambda_{\rm RR}$, analogous to $\Lambda_{\rm QCD}$ for strong interaction. 
In sect.~\ref{sect:sol} we show that the dynamical emergence of  this mass scale can have remarkable consequences on several aspects of the ``cosmological constant" problem. An important role in the picture that we will develop is played by the conformal mode of the metric. Then, 
in sect.~\ref{sect:conf} we examine in more detail the dynamics of the conformal mode in $R^2$ extensions of GR. Finally,
in sect.~\ref{sect:anomaly} we will consider  Einstein gravity, without $R^2$ terms, and the possibility that similar effects arise in this case from the higher-derivative term provided by the anomaly-induced effective action. In sect.~\ref{sect:concl} we summarize our results.

\section{Non-local loop corrections in $R^2$ gravity}\label{sect:one}

\subsection{Notations and conventions}

We consider the theory with action 
\bees\label{SEH}
S&=&S_{\rm EH}+S_{\rm HD}\nn\\
&=& \int d^4x \sqrt{-g}\, \[ \frac{\mplr^2}{2}(R-2\Lambda)-  \(a_1C^2+a_2R^2+a_3 E\)\]\, ,\label{SHD}
\ees
where $\mplr=1/(8\pi G)^{1/2}$ is the (reduced) Planck mass,  $S_{\rm EH}$ 
is the Einstein-Hilbert (EH) action with a cosmological constant, and $S_{\rm HD}$ is the higher-derivative term.
Here 
\be
C^2=\Rmnrs^2-2\Rmn^2+(1/3)R^2\, 
\ee 
is the square of the Weyl tensor, and
\be\label{defE}
E=\Rmnrs^2-4\Rmn^2+R^2\, 
\ee 
is the  Gauss-Bonnet term. Observe that $a_1$, $a_2$ and $a_3$ are dimensionless. We use the MTW sign conventions \cite{MTW}, so in particular $\emn=(-,+,+,+)$. The overall minus sign in front of the higher-derivative terms in \eq{SHD} is part of the  definition of the $a_i$ coefficients, and is chosen so that these terms will appear with a plus sign in the euclidean action, see \eq{Seucl} below. Loop computations are indeed  typically performed with euclidean signature, and it will be important for us to be careful about the signs in the passage from Minkowskian to euclidean signature. The rotation to euclidean signature $(+,+,+,+)$ is obtained performing the  Wick rotation, i.e. introducing euclidean time $t_{\rm E}$ from  $t_{\rm E}=it$. Then $d^4x\ra -i(d^4x)_{\rm E}$, while $\sqrt{-g}\ra g^{1/2}$ and $R$ transforms into the Ricci curvature computed with the euclidean metric, $R_{\rm E}$, without extra minus signs, and similarly for the Riemann and Ricci tensors.  
Then, 
\be
(S_{\rm EH}+S_{\rm HD})_{\rm Mink}=-i \int (d^4x)_{\rm E} \,g^{1/2}\,\[ \frac{\mplr^2}{2} (R_{\rm E}-2\Lambda)
-\(a_1C_{\rm E}^2+a_2R_{\rm E}^2+a_3 E_{\rm E}\)\]\, .
\ee
Defining the euclidean action $S_{\rm Eucl}$ from 
$e^{iS_{\rm Mink}}=e^{-S_{\rm Eucl}}$, i.e. $S_{\rm Eucl}=-iS_{\rm Mink}$, we therefore find
\be\label{Seucl}
S_{\rm Eucl}=\int (d^4x)_{\rm E}\, g^{1/2}\,\[ -\frac{\mplr^2}{2} (R_{\rm E}-2\Lambda)+a_1C_{\rm E}^2+a_2R_{\rm E}^2+a_3 E_{\rm E}\]\, .
\ee
We will henceforth drop the subscript $E$ from euclidean quantities. Whether an action is written in Minkowskian or euclidean signature can be understood from the presence of the factor $\sqrt{-g}$ or $g^{1/2}$, respectively.
We will generically denote by $R^2$ gravity  the theory containing the Einstein-Hilbert action plus all the  terms $R^2$, $C^2$ and $E$. 

\subsection{$R^2$ gravity at low energies}\label{4sect:npert}

Let us first consider $R^2$ gravity in an effective field theory approach, valid in the limit $E\ll \mpl$. In this case the ${\cal O}(R^2)$ terms are simply seen as the first corrections that we meet as we approach the Planck scale.  More precisely, if we linearize over flat space, $\gmn=\emn+\hmn$, the Lagrangian density reads,
schematically (i.e. displaying only the powers of $\hmn$ and of derivatives, without writing explicitly any tensor structure nor numerical factors), 
\be\label{LEHLHD}
{\cal L}_{\rm EH}+{\cal L}_{\rm HD}\sim \mplr^2 (h\pa^2 h +h\pa h\pa h+\ldots)
-a \[ (\Box h)^2 +h  (\Box h)^2 +\ldots \]
\ee
where, for the purpose of this schematic counting,  $a$ denotes generically $a_1$ or $a_2$ (the Gauss-Bonnet term is a topological invariant and we will neglect it for most of the following discussion). In an effective field theory approach the degrees of freedom are read from the Einstein-Hilbert term, so in the spectrum we only have the massless graviton. Canonical normalization is also performed with respect to the Einstein-Hilbert term, i.e. rescaling $\hmn \ra \hmn/\mplr$, and therefore the Lagrangian density is rewritten as
\be\label{schemL}
{\cal L}_{\rm EH}+{\cal L}_{\rm HD}\sim \( h\pa^2 h +\frac{1}{\mplr} h\pa h\pa h+\ldots\)
-\frac{a}{\mplr^2} \[ (\Box h)^2 +\frac{1}{\mplr} h  (\Box h)^2 +\ldots \]\, .
\ee
We see that the contribution to any process from the higher-derivative terms is suppressed, compared to the contributions coming from to the EH action, by a factor of order $aE^2/\mplr^2$. The standard lore is therefore that these corrections 
are irrelevant in the infrared, where $E\ll\mplr$.  

The idea that we want to explore in this paper is that, in an effective action that derives from a  full quantum theory of gravity,  $a_1$ and $a_2$ are dimensionless coupling constants that run under renormalization group, so  
$a_i=a_i(E)$. Depending on the sign of their beta functions, their absolute values can either grow or decrease as we run toward the IR. If a coupling grows in the IR, i.e. if the sign of its beta function corresponds to asymptotic freedom in the UV, we  meet a situation conceptually similar to what happens in QCD. Running toward the IR this coupling will enter a strong coupling regime. The energy scale at which this happens defines a dynamically generated scale, analogous to $\Lambda_{\rm QCD}$. At energies below this new scale,  the functions $a_i(E)$ have a behavior which has nothing to do with the one computed in perturbation theory, and  can in principle even develop singularities as $E\ra 0$. In particular, if in this regime $a_2(E)\propto E^{-4}$,  in coordinate space we would have in the action a term  $Ra_2(\Box)R\propto R\Box^{-2}R$ (see sect.~\ref{sect:NLoneloop} below), which would indeed reproduce the model (\ref{RR}). 

In this section we investigate this scenario. We will first discuss  the running of the couplings  associated to the ${\cal O}(R^2)$ terms, and we will then give arguments, drawing  on the known behavior of singularities in the QCD Green's functions, as well as on the dynamics of the conformal mode in $R^2$ gravity, that suggest that a IR behavior $a_2(\Box)\propto \Box^{-2}$ is indeed  plausible.  
We begin  by reviewing the known results  on perturbative loop corrections to the $a_i$ couplings. 

\subsection{Non-local form factors at one loop}\label{sect:NLoneloop}

In general in QFT the  running of coupling constants, which is more commonly expressed in momentum space, can also be expressed in terms of non-local form factors in the effective action. For instance, at the one-loop level the running of the electric charge in QED can be described in coordinate space by the effective action~\cite{Barvinsky:1987uw} (see also \cite{Dalvit:1994gf,Lombardo:1996gp})
\be\label{qed}
S_{\rm eff}=-\frac{1}{4}\int d^4x\, F_{\mu\nu}\frac{1}{e^2(\Box)}F^{\mu\nu}\, ,
\ee
where
\be
\frac{1}{e^2(\Box)}=\frac{1}{e^2(\mu)}-\beta_0\log\(\frac{-\Box}{\mu^2}\)\, .
\ee
Here $\mu$ is the renormalization scale, $e(\mu)$ is the renormalized charge at the scale $\mu$  and, for a single massless fermion, $\beta_0=1/(12\pi^2)$. The logarithm of the d'Alembertian can be defined for instance from 
\be
\log\(\frac{-\Box}{\mu^2}\)=\int_0^{\infty}dm^2\, \[\frac{1}{m^2+\mu^2}-
\frac{1}{m^2-\Box}\]\, .
\ee
Such non-local actions generate non-local effective
equations of motion  for the expectation values of the quantum fields. The corresponding equations of motion for the in-out matrix elements depend on the Feynman propagator and are acausal, but the effective classical equations of motion for the in-in expectation values, which can be computed using the Schwinger-Keldysh formalism, involve the retarded propagator and are therefore non-local but causal\cite{Jordan:1986ug,Calzetta:1986ey}. In practice, using the in-in formalism turns out to be equivalent to performing the computation in euclidean space, and in the end rotating to Minkowskian signature while at the same time replacing the euclidean $\iBox$ with the Minkowskian $\iBox$ computed with the retarded Green's function, as discussed in \cite{Barvinsky:1987uw} and recently verified explicitly, for the case of $R^2$ gravity, in \cite{Donoghue:2014yha}.

Non-local terms also appear in the renormalization of gravity with higher derivatives, as was first  revealed using the heat-kernel technique in \cite{Barvinsky:1985an, Barvinsky:1987uw}. The result has also been  checked with standard Feynman diagram computations~\cite{Gorbar:2002pw,Gorbar:2003yt}. These non-local terms appear  when we consider gravity semiclassically and quantize massless scalars, massless spinors or massless vector fields over a fixed curved background. Furthermore, they also appear when we quantize gravity itself, as a result of the contribution of gravitons to the quantum loops. The contribution from matter field is very well understood, and summarized in textbooks such as~\cite{Birrell:1982ix,Buchbinder:1992rb}. In contrast, a consistent computation of the running due to graviton loops is a much more subtle issue. We examine these contributions separately.

\subsubsection{Loop corrections from matter fields}

To lowest perturbative order, the loop corrections induced by massless matter fields can be computed in semiclassical gravity, i.e. on a fixed curved background. 
Using $\Rmnrs^2$, $\Rmn^2$ and $R^2$ as  independent terms, the non-local correction to the euclidean action at one-loop takes the form\footnote{We follow the notation in \cite{Donoghue:2014yha}, except that we define the constants $\a,\b,\g$ extracting an overall factor $1/[2(4\pi)^2]$ in front of $S_{\rm NL}^{\rm 1-loop}$, which makes easier the comparison with most of the literature, e.g. with  refs.~\cite{Gorbar:2002pw,Gorbar:2003yt}.}
\bees
S_{\rm NL}^{\rm 1-loop}&=&\frac{1}{2(4\pi)^2}\, \int d^4x\, g^{1/2}\, \[\alpha \,R\log\(\frac{\Box}{\mu^2}\)R+
\beta \Rmn\log\(\frac{\Box}{\mu^2}\)\RMN\right. \nn\\
&&\hspace{3.4cm}\left.+\gamma\, \Rmnrs\log\(\frac{\Box}{\mu^2}\)\RMNRS\]\, ,
\ees
where $\Box$ is the  generally covariant d'Alembertian in euclidean space. Switching to the basis of $R^2,C^2,E$,
this expression can be rewritten as
\bees
S_{\rm NL}^{\rm 1-loop}&=&\frac{1}{2(4\pi)^2}\, \int d^4x\, g^{1/2}\, \[\bar{\alpha} \,R\log\(\frac{\Box}{\mu^2}\)R+
\bar{\beta}\, C_{\mu\nu\rho\sigma}\log\(\frac{\Box}{\mu^2}\)C^{\mu\nu\rho\sigma}\right. \nn\\
&&\hspace*{-1.4cm}\left.+\bar{\gamma}
\(  \Rmnrs\log\(\frac{\Box}{\mu^2}\)\RMNRS -4  \Rmn\log\(\frac{\Box}{\mu^2}\)\RMN
+R\log\(\frac{\Box}{\mu^2}\)R \)
\]\, ,\label{4SNL1loop}
\ees
where $C_{\mu\nu\rho\sigma}$ is the Weyl tensor, while the last structure corresponds to the Gauss-Bonnet term (\ref{defE}).
The relation between the coefficients is \cite{Donoghue:2014yha}
$
\a=\bar{\a}+\frac{\bar{\b}}{3}+\bar{\gamma}$, 
$\beta=-2\bar{\beta}-4\bar{\gamma}$,
$\gamma=\bar{\beta}+\bar{\gamma}$.
The  value of the coefficients 
$\{\bar{\alpha},\bar{\beta}, \bar{\gamma}\}$  can be found e.g. in \cite{Birrell:1982ix} (since the logarithmic terms are fixed by the $1/\eps$ divergences of the effective action). In Table~\ref{tab1} we collect the values of $\{\bar{\alpha},\bar{\beta}, \bar{\gamma}\}$  due to a massless scalar with a generic non-minimal coupling $\xi$ to the curvature, described by the euclidean action
\be
S_s=\frac{1}{2} \int d^4x\, g^{1/2}\, \(\gMN\pam\phi\pan\phi+\xi R\phi^2\)\, ,
\ee
as well as the contributions from massless spinors and  massless vector fields. The value of the scalar field is given in terms of the parameter $\bar{\xi}=\xi-(1/6)$, and $\bar{\xi}=0$ corresponds to a conformally coupled scalar field. 

\begin{table}[t]
\begin{center}
\begin{tabular}{|l|c|c|c|} 
\hline 
\phantom{$\[ \frac{(A^2)}{B}\]$}            &$\bar{\alpha}$         & $\bar{\beta}$   &$  \bar{\gamma}$\\
             \hline
scalar\phantom{$\[ \frac{(A^2)}{B}\]$}       &$(1/2) \bar{\xi}^2$ &   1/120             & $-1/360$\\ \hline
fermion                                                           & 0                              &  6/120             & $-11/360$\\ \hline
vector                                                              & 0                             &  12/120           & $-62/360$\\ \hline
\end{tabular}  
\end{center}
\caption{The value of the parameters $\{\bar{\alpha},\bar{\beta}, \bar{\gamma}\}$ due to loops of   massless matter fields. 
\label{tab1}}
\end{table}

The cosmological effect of these non-local terms has  been recently studied in \cite{Donoghue:2014yha}. Even if enhanced by a logarithmic factor, a term $R\log(-\Box/\mu^2)R$ can compete with the Einstein-Hilbert term $\mplr^2R$ only for nearly Planckian curvatures. Therefore, the inclusion of the UV form of the loop corrections can be relevant for addressing issues such as the resolution of cosmological singularities near the big-bang, or the emergence of the classical behavior from the full quantum theory, which are the issues addressed in  \cite{Donoghue:2014yha}.\footnote{See also \cite{Rinaldi:2014gha}, where one-loop corrections to the $R^2$ theory computed expanding over de~Sitter space, and involving corrections of the form  $R^2\log R/\mu^2$, are used to construct an early-universe  inflationary model. Non-local extension of the $R^2$ theory relevant in the UV have also  been studied, with different motivations, in several papers, see e.g. \cite{Biswas:2011ar,Modesto:2011kw}.}
Here we are rather interested in the opposite limit  of curvatures much smaller than $\mplr$ and  very large distances, i.e. in the far IR, with the aim of understanding the origin of dark energy.

It is also  instructive to compute the one-loop renormalization of the $R^2$ and $C^2$ terms induced by massive (rather than massless) matter fields in a curved background, to understand the interplay between non-locality  and the  mass of the field. This computation has been 
performed  in \cite{Gorbar:2002pw,Gorbar:2003yt} for scalar, spinor and vector fields (neglecting the renormalization of the Gauss-Bonnet term), using a mass-dependent renormalization scheme that allows one to correctly recover the decoupling of massive fields in the limit of small momenta. 
Consider for instance a real scalar field  with mass $m_s$ and a generic non-minimal coupling $\xi$, described  by the euclidean action
\be
S_s=\frac{1}{2} \int d^4x\, g^{1/2}\, \(\gMN\pam\phi\pan\phi+m_s^2\phi^2+\xi R\phi^2\)\, .
\ee
The corresponding one-loop contribution to the euclidean effective action 
is \cite{Gorbar:2002pw,Gorbar:2003yt}\footnote{This result was  first obtained in \cite{Gorbar:2002pw}  using 
a non-covariant computation based on an expansion of the action over flat space, and it was then checked in the same paper using the covariant heat-kernel technique.  In the covariant technique the quantity which is computed is $\bar{\Gamma}^{(1)}$, defined by 
$\det^{-1/2}( -\Box+m^2+\xi R) =\exp\{\bar{\Gamma}^{(1)}\}$. Since the euclidean action enters as $e^{-S_{\rm Eucl}}$, we have $S_{\rm Eucl}^{\rm 1-loop}=-  \bar{\Gamma}^{(1)}$. This is the origin of the overall sign difference between the result of the covariant computation (eqs.~(2.15) and (2.18) of  \cite{Gorbar:2002pw}) and of the non-covariant computation, eq.~(3.9) of ref.~\cite{Gorbar:2002pw}. I thank E.~Gorbar for clarifying me this point.\label{foot:Gamma}}
\bees
S^{\rm 1-loop} &=& -\frac{1}{2(4\pi)^2}\,\int d^4x
\,g^{1/2}\, \Big\{\,\frac{m_s^4}{2}\Big(\frac{1}{\eps}
+\frac{3}{2} \Big) \,+ \,\bar{\xi}m_s^2R\,
\Big(\frac{1}{\eps}+1\Big) \nonumber\\
&+& \frac{1}{2}\,C_{\mu\nu\rho\sigma} \,\Big[\frac{1}{60\,\eps}+k_W(\Box)\Big]
C^{\mu\nu\rho\sigma} \,+
\,R\,\Big[\,\frac{1}{2\eps}\,\bar{\xi}^2\, +
k_R(\Box)\,\Big]\,R\,\Big\}\, ,\label{ShapSola29}
\ees
where $1/\eps=2/(4-D)+\log (4\pi\mu^2/m_s^2)-\gamma_{\rm E}$ comes from the dimensional regularization scheme in $D$ space-time dimensions,  and the form factors $k_W(\Box)$ and $k_R(\Box)$ are given by
\bees
k_W(\Box) &=& \frac{8A}{15\,a^4}
\,+\,\frac{2}{45\,a^2}\,+\,\frac{1}{150}\,, \label{kR}\\
k_R(\Box) &=&
\bar{\xi}^2A
+ \(\frac{2A}{3a^2}-\frac{A}{6}+ \frac{1}{18}\)\bar{\xi}
+ A\( \frac{1}{9a^4}- \frac{1}{18a^2}
+ \frac{1}{144} \)\nn\\
&&+ \frac{1}{108\,a^2} -\frac{7}{2160}
\,, \label{kW}
\ees
where
\be
A\,=\,1-\frac{1}{a}\log\,\Big(\frac{2+a}{2-a}\Big)\,, \qquad a^2 =
\frac{4\Box}{\Box - 4m_s^2}\, .
\ee
A few comments on this result are in order. First, observe that the first two terms in \eq{ShapSola29} correspond to the renormalization of the cosmological constant and of Newton constant, respectively. Even if they depend on the renormalization scale $\mu$ (through the parameter $\eps$), they have no dependence on the $\Box$ operator and, in this sense, no running in momentum space. This is a trivial consequence of the fact that $\Box\Lambda=0$, while $\Box R$ is a total derivative, so we cannot  obtain a form factor by acting with the $\Box$ operator on $\Lambda$ or on $R$. In contrast, the 
$C_{\mu\nu\rho\sigma}^2$ and the $R^2$ terms acquire corrections of the form
$C_{\mu\nu\rho\sigma} k_W(\Box)C^{\mu\nu\rho\sigma} $ and
$Rk_R(\Box)R$, in full analogy with \eq{qed}.
In the UV limit $\Box/m_s^2\gg 1$ we have (see also \cite{Avramidi:1989er,Dalvit:1994gf})
\bees
k_W(\Box)&\simeq &-\frac{1}{60}\log\frac{\Box}{m_s^2}+ \frac{23}{450} 
+{\cal O}\(\frac{m_s^2}{\Box}\log\frac{m_s^2}{\Box} \)
\, ,\label{kWUV}\\
k_R(\Box)&\simeq &-\frac{1}{2}\bar{\xi}^2\log\frac{\Box}{m_s^2}+
\(-\frac{1}{1080}+\frac{\bar{\xi}}{18}+\bar{\xi}^2\)
 +{\cal O}\(\frac{m_s^2}{\Box}\log\frac{m_s^2}{\Box} \)
\, .\label{kRUV}
\ees
Therefore, after subtracting the divergent parts, in the limit $\Box/m_s^2\gg 1$ we remain with
\be
S^{\rm 1-loop} = \frac{1}{2(4\pi)^2}\,\int d^4x\,g^{1/2}\,
\[ \frac{1}{2}\bar{\xi}^2 \,R\log\(\frac{\Box}{m_s^2}\)R+
\frac{1}{120}\, C_{\mu\nu\rho\sigma}\log\(\frac{\Box}{m_s^2}\)C^{\mu\nu\rho\sigma}\]\, ,
\ee
(plus local terms independent of $\Box$) in full agreement with \eq{4SNL1loop} and Table~\ref{tab1} (the renormalization of the Gauss-Bonnet term was not included in this computation).
Observe also that, in the conformal case  $m_s=0, \bar{\xi}=0$, the form factor $k_R(\Box)$ becomes local, $k_R(\Box)=-1/1080$, in agreement with the prediction that can be obtained directly from the conformal anomaly.\footnote{Again, checking the sign is a bit tricky due to the many different conventions in the literature.  The standard anomaly result can be read from  eq.~6.124 and Table 1 on page  179 of the Birrell and Davies textbook ~\cite{Birrell:1982ix}. One must however take care that Birrell and Davies use the signature $(+,-,-,-)$ and their definition of Riemann tensor is opposite to MTW. To go to MTW sign conventions we therefore must switch $\gmn \ra -\gmn$ (so $\Box \ra -\Box$) and $R \ra +R$.
Taking into account that this flips also the sign of the trace $T^{\mu}_{\mu}$, we find that the correct result for the anomaly of a single massless conformally coupled scalar field, in MTW conventions, is
$ T = + \frac{1}{180 (4\pi^2)} \Box R$. From the signature $(-,+,+,+)$ we can then trivially rotate to euclidean space without getting extra minus signs. In  eq.~(3.12) of \cite{Gorbar:2002pw} the sign of the anomaly in the euclidean space was incorrectly taken to be a minus, and the result was checked using $\bar{\Gamma}^{(1)}$ instead of $S_{\rm Eucl}^{\rm 1-loop}=-  \bar{\Gamma}^{(1)}$, see footnote~\ref{foot:Gamma}, so these two sign errors compensated.}
 
In contrast,  in the opposite limit $\Box/m^2\ll 1$, \eqs{kR}{kW} give
\be\label{dec}
k_W(\Box),k_R(\Box)={\cal O}(\Box/m^2)\, , 
\ee
corresponding to the decoupling of particles with mass large compared to the momentum scale, which is explicit in the mass-dependent subtraction scheme used in \cite{Gorbar:2002pw,Gorbar:2003yt} (see also the discussion in \cite{Manohar:1996cq}).
Furthermore, beside being small, a term ${\cal O}(\Box/m^2)$ is local. This shows explicitly that non-local terms appear in the effective action as a consequence of the running in the loops of particles that are either massless, or light with respect to the energy scale considered.

\subsubsection{Gravitational loop corrections in Einstein gravity}

In semiclassical gravity, $a_1,a_2,a_3$ are simply parameters of the effective action. 
When we go beyond the semiclassical approximation, and treat gravity itself at the quantum level, $a_1$ and $a_2$  become genuine coupling constants, and the perturbative expansion is organized in terms of the three coupling of the theory $G, a_1$ and $a_2$ (while  the Gauss-Bonnet term is a topological invariant, and will be neglected in the following). 
Their renormalization group equations also receive contributions from the purely gravitational sector, i.e. from graviton loops. However, these contributions depends crucially on the UV completion of the theory, as we now review.  

If one starts from the Einstein-Hilbert action without higher-derivative terms and computes the one-loop corrections, one in principle finds $R^2$ and $C^2$ terms, as was first shown in the classic paper by 't~Hooft and Veltman \cite{'tHooft:1974bx}. The original computation of \cite{'tHooft:1974bx} gives, in our notation, $\bar{\a}=90/360=1/4$ and $\bar{\beta}=126/360=7/20$, which are the values quoted in~\cite{Donoghue:2014yha}. However, these divergences have no physical meaning, since they  can be set to zero using the equations of motion~\cite{'tHooft:1974bx}, so that pure gravity at one loop is finite on-shell. Equivalently, they can be set to zero with a field redefinition
$\gmn\ra \gmn+\eps (c_1\Rmn+c_2 R\gmn)$. More generally, if one quantizes the pure Einstein-Hilbert action, the coefficients of the $R^2$ and $\Rmn^2$ terms which are generated at one loop depend on the gauge used, and one can find gauges in which these divergences are absent, so that the theory is one-loop finite even off-shell~\cite{Kallosh:1978wt}. Thus, in pure Einstein-Hilbert gravity there is no sense in which we can compute the gravitational contribution to the running of the coupling constants associated to the $R^2$ terms. The same is true if we  consider the  higher derivative theory (\ref{SHD}) in an effective action approach, since the higher-derivative term add extra vertices, but does not change the structure of the propagator.

\subsubsection{Gravitational loop corrections in Stelle theory}\label{sect:Stelleth}

The situation  changes in an interesting manner if, rather than using an effective action approach, in which the higher-derivative term is seen as a correction valid only in the low-energy limit, we try to take the action (\ref{SHD})
as a full UV complete theory, and quantize it. In this approach, the higher-derivative theory (\ref{SHD}) is usually known as Stelle theory.  In this case the propagator is read from the full quadratic term in \eq{LEHLHD} and, again schematically, is of the form 
\be\label{Gdik}
G(k)\sim \frac{1}{\mplr^2k^2+ak^4}\, ,
\ee
(before any canonical normalization) so that now it goes as $1/k^4$ in the far UV. This improved behavior of the propagator allowed Stelle to prove that this theory is renormalizable~\cite{Stelle:1976gc}.\footnote{Of course, the full story is  more complex. The $R^2$ and $C^2$ terms also lead to vertices growing as $k^4$, rather than  $k^2$ as in Einstein-Hilbert gravity. In GR, since the propagator go as $1/k^2$ and the vertices as $k^2$, the superficial degree of divergence is $D=4L-2I+2V$, where $L$ is the number of loops, $I$ of internal lines, and $V$ of vertices. Combined with the topological relation $L=1+I-V$ this gives $D=2L+2$, so the degree of divergence grows with the number of loops. In contrast, for ${\cal L}_{\rm HD}$, $D=4L-4I+4V$ which, together with 
$L=1+I-V$, gives $D=4$, independent of the number of loops~\cite{Stelle:1976gc}. This improved behavior is at the basis of Stelle's proof of renormalizability. 
Observe also that, for  renormalizability, it is crucial to include both the $R^2$ and the $C^2$ terms in the action, since they contribute separately to the propagator of the spin-0 and massive spin-2 excitation of the theory. Otherwise,  the propagator of one of these excitations would still go as $1/k^2$, while the vertices grow as $k^4$, leading to an even worse UV behavior than in Einstein-Hilbert gravity. 
Furthermore, even the Gauss-Bonnet term is required for multiplicative renormalizability (see the recent discussion in ~\cite{Einhorn:2014gfa}).}
Having a renormalizable quantum theory of gravity is quite  remarkable. This however  comes at a very high price. Namely, the spectrum of $R^2$ gravity, which is now read from the full propagator, consists of the usual massless graviton, 
a massive spin-2 particle and a massless spin-0 particle~\cite{Stelle:1977ry,Fradkin:1981iu}.
The massive spin-2  state has negative kinetic energy, so is a ghost, and its squared mass is
\be\label{m22}
m_2^2=\frac{\mplr^2}{2 a_1}\, .
\ee
The massive spin-0  state has a normal kinetic term, and squared mass
\be\label{m02}
m_0^2=-\frac{\mplr^2}{12 a_2}\, ,
\ee
and is a tachyon if $a_2>0$. Because of the ghost, and (in case) of the tachyon, $R^2$ gravity, treated non-perturbatively as a full UV complete theory, is not a consistent theory  (otherwise, the problem of finding a consistent theory of quantum gravity would have been solved in the late 1970s...). There have been attempts at exorcising the ghost based on the idea that radiative corrections shift its pole into the complex plane~\cite{Tomboulis:1977jk,Antoniadis:1986tu}, so that the ghost becomes unstable and does not appear in the asymptotic states. To leading order in $1/N$, where $N$ is the number of conformally coupled matter fields, the theory turns out indeed to be unitary. However, attempts at proving unitarity beyond leading order based on the gauge-dependence of the position of the ghost poles fail~\cite{Johnston:1987ue}, so what happens  to all orders is basically unknown. In any case,
there is a long tradition of ignoring the ghost problem and see what the $R^2$ theory has to say at a full non-perturbative level, in the hope that this will at least help to unveil properties of a fully consistent theory of quantum gravity. In particular, one can compute the running of the coupling constants associated to the ${\cal O}(R^2)$ terms. As we mentioned above, in Einstein-Hilbert gravity this computation leads to beta functions that are not even gauge-independent.
In contrast, in Stelle theory the computation is well-defined, and the  beta functions  associated to the $R^2$, $C^2$  and Gauss-Bonnet terms are physical and gauge-independent (while, for the Newton constant and the cosmological constant, only the dimensionless combination 
$G\Lambda$  has a gauge-independent beta function)~\cite{Julve:1978xn,Fradkin:1981iu,Avramidi:1985ki}. For Stelle theory,
the  computation of the gravitational contribution to the beta functions of the $a_i$ coupling was first performed correctly in \cite{Avramidi:1985ki} (building on earlier work in  \cite{Fradkin:1981iu,Julve:1978xn}), and the result was  confirmed in \cite{deBerredoPeixoto:2004if}.  The result is conveniently expressed trading $a_1$ and $a_2$ for two couplings $\lambda$ and $u$ defined by
\be\label{a1a2lu}
a_1=\frac{1}{\lambda}\, ,\qquad a_2=-\frac{u}{3\lambda}\, ,
\ee
so that the Minkowskian action (\ref{SEH}) reads (neglecting the Gauss-Bonnet term and the cosmological constant)
\be
S=\int d^4x \sqrt{-g}\, \[ \frac{\mplr^2}{2}R-  \frac{1}{\lambda} C^2+ \frac{u}{3\lambda}R^2\]\, .
\ee
The corresponding  beta functions
are given in eqs.~(19)-(24) of \cite{Avramidi:1985ki}. In particular, introducing the logarithm of the energy-scale parameter $t=(4\pi)^{-2}\log E/\mu_0$, the purely gravitational contribution is
\bees
\frac{d\lambda^{-1}}{dt}&=&\bar{\beta}_g\, ,\label{betal}\\
\frac{du}{dt}&=& -\lambda \( \frac{10}{3}u^2 +\frac{183}{10}u+\frac{5}{12}\)\, ,\label{betau}
\ees
where $\bar{\beta}_g=133/10$ is positive.  It is important to stress that  these beta functions have been computed in the UV limit $E\gg\mplr$. In the  regime $E\ll\mplr$ which is interesting for us, these equations will receive corrections from the Einstein-Hilbert term, which have not been computed.
The solution of \eq{betal} is
$\lambda^{-1}(t)=\lambda^{-1}(t=0)+\beta_g t$,
i.e.
\be\label{ldiE}
\lambda(E)=\frac{\lambda(\mu)}{1+\lambda(\mu) \beta_g\log (E/\mu)  }\, .
\ee
where $\beta_g=\bar{\beta}_g/(4\pi)^2$. Whether this corresponds to a function $\lambda(E)$ that becomes large in the UV and small in the IR, or viceversa,  depends of course of the sign of $\lambda(\mu)$. Since $\bar{\beta}_g>0$, 
if $\lambda(\mu)>0$ we have  $\lambda(E)\ra 0$ as $E\ra\infty$, and conversely $\lambda(E)$ gets large and formally diverges at a finite energy scale  in the IR.
The opposite situation takes place if $\lambda(\mu)<0$.  Observe from
Table~\ref{tab1} that the contribution to the beta function of $a_1=\lambda^{-1}$ coming from each
types of matter fields has the same sign as the gravitational contribution, so the same behavior takes place in the full theory with gravitational plus matter loops.

The evolution of $u$ is instead determined, at least in the super-Planckian regime in which \eq{betau} holds, by the two roots of the 
equation $(10/3)u^2 +(183/10)u+(5/12)=0$, which are given by $u_1\simeq -5.46714$, $u_2\simeq -0.02286$. Writing 
\eq{betau} as 
\be
\frac{du}{dt}=-\frac{10}{3}\lambda (u-u_1)(u-u_2)\, ,
\ee
we can easily read the sign of $du/dt$ and the direction of the RG flow, for different values of $u$, and a given sign of $\lambda$. For $\lambda >0$ the one-dimensional flow in the $u$ variable is depicted in the upper panel of Fig.~\ref{fig:flow_u} and for $\lambda <0$ is shown in
the lower panel. Observe in particular that, for $\lambda>0$, $u=u_2$ is a UV fixed point, which attracts all initial values
$u(t=0)>u_1$. For $\lambda<0$ the value  $u=u_1$ is a UV fixed point at its attraction basin is given by $u(t=0)<u_2$.

\begin{figure}[t]
\centering
\includegraphics[width=0.5\columnwidth]{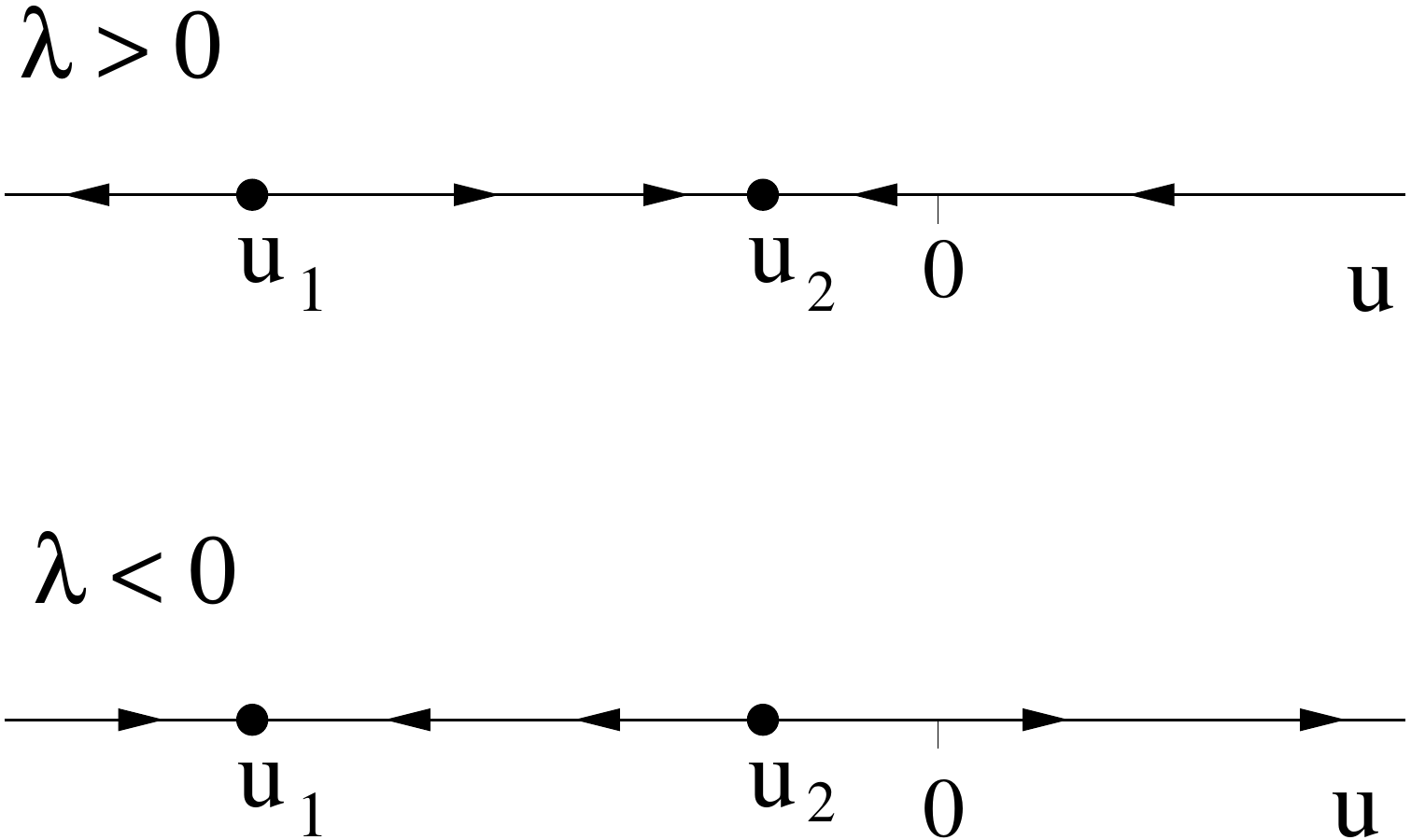}
\caption{\label{fig:flow_u} The RG flow of $u(t)$ for $\lambda>0$ (upper panel) and  $\lambda<0$ (lower panel). The direction of the arrows indicate the flow of $u$ as $t$ increases, i.e. as we move from the IR to the UV. For a given initial value in the UV, the flow as we run toward the IR regime is therefore obtained reversing the sign of the arrows.
\label{fig:flow_u}
}
\end{figure}

\begin{figure}
\begin{center}
\begin{tikzpicture}[line width=1 pt, scale=2]
\draw[graviton] (1,0)--(2,0);
\node at (2.5,0) {+};
\draw[graviton] (3,0)--(4.02,0);
\draw[graviton] (5,0) arc (0:360:0.5);
\draw[graviton] (5,0)--(6,0);
\end{tikzpicture}
\caption{The graviton propagator and a one-loop correction. \label{fig:loop}}
\end{center}
\end{figure}
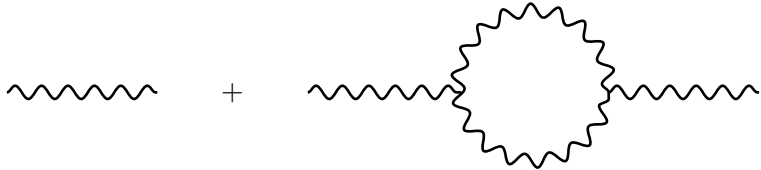

One might wonder how is it possible that a computation which is well-defined in the full $R^2$  theory is not well-defined in Einstein-Hilbert gravity, since the latter is formally obtained taking the limit $a_1,a_2\ra 0$ in the action (\ref{Seucl}).
The reason is that, when we take the $R^2$ theory as a full non-perturbative theory,  the structure of the divergences, which determine the beta functions, changes discontinuously if we set $a_1$ or $a_2$ to zero. Indeed, consider for instance a loop correction to the graviton propagator such as that shown in Fig.~\ref{fig:loop}. Even when the momentum $p$ of the external line is sub-Planckian, $|p^2|\ll\mplr^2$, in the loops we will have propagators such as $G(k)$, given in \eq{Gdik}, and $G(k-p)$, where $k$ is the loop integration variable and $p$ the external momentum. The divergence is determined by the behavior of the propagator (and of the vertices) for $k\ra \infty$. For any $a\neq 0$ the propagator at $k\ra\infty$ behaves as $1/(ak^4)$, even when the external momenta are sub-Planckian, while for $a=0$ it behaves as $1/k^2$ and the structure of the divergences is completely different. In other words, the limit  $a\ra 0$   does not commute with the integrations over $d^4k$ in the loop integrals. 
So, despite the fact that $S_{\rm HD}$ is a higher-derivative term, which naively should be irrelevant at low energies, the structure of the divergences of Stelle theory is completely different from that of GR, even when the external lines are sub-Planckian, and the limit $a\ra 0$ (or, more precisely $a_1\ra 0$ or $a_2\ra 0$) of the beta functions of $R^2$ gravity  is not smooth. This fact has already been stressed in sect.~3.1 of \cite{Fradkin:1981iu}, where in particular it was observed that
the beta function of the pure Weyl theory cannot be obtained setting $a_2=0$ in the beta function of the theory with a higher-derivative term $a_1C^2+a_2R^2$.

\subsubsection{Gravitational running in a consistent theory of quantum gravity}

The result (\ref{betal}, \ref{betau}) cannot be applied directly to our problem, namely the running of $a_2$ at low energies in the actual universe, for two  reasons. First, the computation leading to \eqs{betal}{betau} has been performed in the deep UV region, $E\gg\mplr$. In the  regime $E\ll\mplr$, which is interesting for us, these equations will receive contributions from the Einstein-Hilbert term, which have not been computed. This is just a technical problem. However, a second and  more fundamental problem is that,
barring a resolution of the ghost problem, Stelle theory can anyhow only be taken as a toy model, rather than a theory describing  Nature.

However what we learn from this toy model is that, in a  theory of gravity with a good UV behavior, the computation of the beta functions associated to the $a_i$ couplings is well defined, and these quantities run with energy. In the end we expect that, whatever will be the consistent fundamental quantum theory of gravity,  in that context it will be possible to compute consistently the running of the coupling constant associated to the $R^2$ and $C^2$ term in the corresponding low-energy effective action, including the gravitational contribution.\footnote{For instance, in string theory one could in principle compute  the string-loop correction to gravitational scattering amplitudes directly at the world-sheet level, and reconstruct from it the corresponding loop corrections in the target-space effective action. Actually, in the effective action of type II string theory in 10 dimension there is no $R^2$ term because of supersymmetry. However, in general, the presence and the structure  of ${\cal O}(R^2)$ corrections that one obtains in four dimensions will depend on the type of string theory considered, as well as on how supersymmetry is broken in the compactification to four dimensions. Observe that it is sometimes argued that, since the fundamental string theory is ghost-free, in the low-energy effective action can appear only the ghost-free combination given by the Gauss-Bonnet term. This argument is however wrong. Being ghost-free is a property of the full theory, and if we write the effective action derived
from a ghost-free theory in an expansion in derivatives and truncate this  expansion to a finite order (e.g. in our case keeping only the terms up to ${\cal O}(R^2)$) apparent ghost state will in general appear, with a mass of the order of the cutoff scale of the effective theory (in our case $\mplr$ or the string mass scale), but they are just an artefact of the truncation. See e.g. \cite{Burgess:2014lwa} for a nice explicit example.}

From the above analysis we also learn that the running of the  couplings $a_i(E)$ depends on the precise form of the UV completion of the theory, even at $E\ll\mplr$. Indeed, the running of $a_i(E)$
computed in a consistent  UV completion of GR cannot reduce at low energies to that  computed in the effective low-energy theory (i.e. in the Einstein-Hilbert action supplemented by the higher-derivative terms, considered as corrections in the effective field theory sense), simply because there is no consistent result in the low-energy effective theory. The beta functions computed in the low-energy effective action based on the Einstein-Hilbert term, in which the $R^2$ terms are treated as corrections, are not even gauge-invariant. This of course does not happens e.g. in QCD, because QCD is renormalizable, which in the end means that, even if we eventually embed it in a larger theory, the details of the UV completion, much as the details of the regularization scheme, are irrelevant at low energies. This is not the case for Einstein-Hilbert gravity, which is not renormalizable.
In a sense, this phenomenon is an interesting example of UV/IR mixing, since it implies that the IR physics of GR can sensitive
to the UV completion.

Thus, the points that we will take home from the following analysis are the following.

\begin{enumerate}

\item In a fundamental and consistent quantum theory of gravity the couplings associated to the $R^2$ and $C^2$ term will in general  run with energy, and their beta functions will  well-defined and gauge-independent.\footnote{Of course, in some special case, the beta functions can be zero, e.g. if they are protected by supersymmetry.} 

\item  The precise form of the corresponding RG equation, even in the IR regime $E\ll\mplr$, depends on the specific UV completion of the theory, and is therefore at present completely unknown. 
We cannot use the results from GR plus higher-derivative terms, considered in an effective action approach, because the corresponding beta functions are non even gauge-invariant. We cannot use the results of Stelle theory (\ref{betal}) and (\ref{betau}), either, because, first of all, these results only hold in the regime $E\gg\mplr$ and, more importantly, Stelle theory  anyhow is not a consistent UV completion of gravity because of the ghost.

\end{enumerate}

It is also interesting to observe that, to obtain the model (\ref{RR}) in the IR, it is not a priori necessary to match the sign of $a_2(\mu)$ in the UV with its (positive) sign in the IR. Indeed, Fig.~\ref{fig:flow_u} gives an explicit example of a RG flow that changes the sign of $u$, and therefore of $a_2$.
A renormalization group flow that changes the sign of the coupling $a_2(\Box)$, flowing from a value $a_2(\mu)<0$ at some scale $\mu$ corresponding to the energy scale of inflation,  to a value $a_2(\Box)=+\Lambda_{\rm RR}^4/\Box^2$ in  the IR (where $\Lambda_{\rm RR}$ is a constant that we will discuss below), would be particularly intriguing since it would interpolate between the Starobinsky model~\cite{Starobinsky:1980te} in the UV and the non-local model (\ref{RR}) in the IR, thereby providing a unified description of inflation in the early universe and dark energy in the present epoch.\footnote{I thank  Eugenio Bianchi for a discussion from which this idea emerged.}

\section{Infrared dynamics}\label{sect:IR}

Even if we do not know the form of the IR running of the couplings $a_i(E)$ in the fundamental theory of gravity, we can still analyze some general scenarios. In the end, there are basically two natural possibilities. Either $|a_2(E)|$ grows in the IR, or it decreases. The interesting situation, from our point of view, takes place when in the IR  $|a_2(E)|$ grows and, at least at sufficiently low energies,  $a_2(E)>0$   (since this sign allows us to eventually match $a_2(\Box)$ with the sign of the model \ref{RR}). In this case, the situation will be quite similar to that in QCD. To stress the similarity, let us use the notation $a_2(E)=g^2(E)$ and assume that $g^2(E)$ is asymptotically free in the UV so, as long as we are in the perturbative regime,
\be\label{gdiE}
g^2(E)=\frac{g^2(\mu)}{1+g^2(\mu) \a_0\log (E/\mu)  }\, ,
\ee
with $\a_0>0$. This implies that $g^2(E)$ eventually becomes large in the IR. Similarly, we can write $a_1(E)=f^2(E)$ and study the scenario in which $f^2(E)$ is asymptotically free in the UV, and
\be\label{gdiE}
f^2(E)=\frac{f^2(\mu)}{1+f^2(\mu) \b_0\log (E/\mu)  }\, ,
\ee
with $\b_0>0$. To understand what happens when  we enter more deeply in the IR regime is of course a  difficult problem.
As already observed recently \cite{Einhorn:2014gfa,Alvarez-Gaume:2015rwa}
a pure $R^2$ theory, with no Einstein-Hilbert term, has a confining Newtonian potential $V(r)\propto r$, because of the $1/|\vk|^4$  behavior of the propagator in the static limit. Its infrared limit is therefore very different from Einstein gravity, and indeed one can even expect   that in this theory gravitons should be confined. When we  also add  the Einstein-Hilbert term, at energies $E\ll\mplr$ the standard $\mplr^2k^2$ term from Einstein-Hilbert action  dominates over the $k^4$ term and we recover the standard behavior of gravity. Still, in the infrared, from the point of view of the quantum theory of gravity, we are in a peculiar situation: loops associated to an expansion in  powers of  Newton's constant are suppressed by powers of $GE^2\sim E^2/\mplr^2\ll 1$ and are therefore irrelevant. However, if the the couplings $a_1(E)$ and/or  $a_2(E)$  run toward a strong coupling regime and formally diverge at some energy in the IR, quantum loops associated to them can be important. As a result, the form factors associated to the $R^2$ and $C^2$ terms can become quite different in the IR. To determine the expression of these form factors  in the IR is of course a difficult non-perturbative problem. However,  experience with QCD allows us to understand some aspects of it. In particular, we know that the logarithmic running  of an asymptotically free coupling constant  generates  by dimensional transmutation  a renormalization group invariant mass scale, that we denote by $\Lambda_{\rm RR}$, analogous to $\Lambda_{\rm QCD}$.\footnote{A similar scenario has been discussed  recently, in
\cite{Einhorn:2014gfa,Salvio:2014soa}, although in these cases dimensional transmutation is rather used to generate the Planck scale from a pure higher-derivative gravity. The idea has been around in different forms since a long time, see e.g.~\cite{Zee:1983mj,Nepomechie:1983yq}.   In ref.~\cite{Tong:2014era} similar ideas have been applied to the coefficient of the Gauss-Bonnet term, using arguments from supergravity.} More precisely, once we fix observationally the value of $f^2(\mu)$  at some scale $\mu$,  formally, as we run toward the IR, the one-loop expression for $f^2(E)$ will hit infinity at some scale
$\Lambda_f$ given by 
\be
\Lambda_f=\mu \exp\left\{-\frac{1}{\beta_0 f^2(\mu)}\right\}\, .
\ee
Similarly, the one-loop expression for $g^2(E)$ will hit infinity at some scale
\be
\Lambda_g=\mu \exp\left\{-\frac{1}{\a_0 g^2(\mu)}\right\}\, .
\ee
As we run toward the IR, the first scale that we hit, where one of the two running couplings formally diverges, defines the strong-coupling scale of the theory, so $\Lambda_{\rm RR}={\rm max}(\Lambda_f,\Lambda_g)$.\footnote{One might wonder whether one of the couplings could still remain weakly coupled in the regime where the other is  strongly coupled. The answer is in general no. For instance, a pure Weyl-square theory, which only has the $f^2$ coupling, is not multiplicatively renormalizable, and generates a $R^2$ term already at the two-loop level~\cite{Fradkin:1983tg}. Therefore, when $f^2$  enters the strong-coupling regime, it necessarily induces a strong coupling in  $g^2$. For this reason, even if for instance the coupling $a_2$ associated to the $R^2$ term should not have the appropriate sign for becoming strongly coupled according to its own one-loop RG equation, it could eventually be driven toward  strong coupling by the coupling $a_1$.}
We therefore have a mass scale that appears in the IR regime by dimensional transmutation.
Then, in the strong coupling region,
$f^2=f^2(\Box/\Lambda^2_{\rm RR})$ and $g^2=g^2(\Box/\Lambda^2_{\rm RR})$. In principle, it is possible to obtain a power-like dependence of 
$f^2$ and $g^2$ on $\Box$, from a resummation of leading logs, of the form\footnote{See \cite{Hamber:2005dw} for similar considerations applied to the running of Newton's constant, and \cite{Taylor:1989tm,Taylor:1989ua} for earlier related ideas on how  the large-distance behavior of  quantum gravity might affect the cosmological constant.}
\be\label{resumnu}
\sum_{n=0}^{\infty}\, \frac{(-\nu)^n}{n!}\, \(\log\frac{-\Box}{\Lambda^2_{\rm RR}}\)^n=
\(\frac{\Lambda^2_{\rm RR}}{-\Box}\)^{\nu}\, ,
\ee
for some value of $\nu$. The model (\ref{RR}) would then be obtained if $\nu=2$. Again, to determine the precise form of $g^2(\Box)$ in the IR is of course  a difficult perturbative problem. However,  it is interesting to observe that in QCD the introduction in the action of a non-local term
\be\label{Fmn2}
-\frac{m^2}{2}\int d^4x\, F_{\mu\nu}^a \[\frac{1}{D^2}\]^{ab}F_{\mu\nu}^b\, ,
\ee
(where  $D_{\mu}^{ab}=\d^{ab}\pam-gf^{abc}A_{\mu}^c$ is the covariant derivative and $m$ is a mass scale) correctly reproduces the results on the non-perturbative gluon propagator in the IR, obtained from operator product expansions and lattice QCD~\cite{Boucaud:2001st,Dudal:2005na,Capri:2005dy,Dudal:2007cw,Dudal:2008sp} (see also the review~\cite{Boucaud:2011ug}). In the gauge $\pam A_{\mu}=0$ we have 
\be
-\frac{m^2}{2}\int d^4x\, F_{\mu\nu}^a \[\frac{1}{D^2}\]^{ab}F_{\mu\nu}^b=
m^2\int d^4x\, A_{\mu}^aA_{\mu}^a +{\cal O}(A^3)\, ,
\ee
and therefore (\ref{Fmn2}) is a mass term for the gluon. Observe that  the expression (\ref{Fmn2}) is  non-local but  gauge-invariant. 
Physically, the introduction of this term in the effective action can  be seen as describing the emergence of a dimension-2 gluon condensate,
$\langle A_{\mu}^aA_{\mu}^a\rangle\neq 0$. The same  operator has been proposed in 3-dimensional YM theory \cite{Jackiw:1995nf,Jackiw:1997jga}, to describe the generation of a mass gap due to IR fluctuations.

The situation is quite similar for the term $R\Box^{-2}R$ in the gravitational case. Indeed, consider the conformal mode $\sigma(x)$ of the metric, 
defined choosing a fixed fiducial metric $\gbmn$ and writing
\be\label{gmnsigma}
\gmn(x)=e^{2\sigma(x)}\gbmn(x)\, .
\ee
Using for simplicity a flat fiducial metric $\gbmn=\emn$, the Ricci scalar computed from the metric $\gmn=e^{2\sigma(x)}\emn$ is
\be
R=-6 e^{-2\sigma}\( \Box\sigma +\pam\sigma\paM\sigma\)\, .
\ee
Therefore, to linear order in $\sigma$,
\be
R=-6\Box\sigma +{\cal O}(\sigma^2)\, ,
\ee
and (upon integration by parts)
\be\label{m2s2}
 R\frac{1}{\Box^2} R=36 \sigma^2 +{\cal O}(\sigma^3)\, .
\ee
The  $\sigma^2$ term is  just a mass term for the conformal mode, while the higher-order interaction terms on the right-hand side of \eq{m2s2} (which are non-local even in $\sigma$) are required to reconstruct a diff-invariant quantity.\footnote{Actually, there is here a subtle  technical point, both in the gravitational and YM case. Indeed,  in \eq{m2s2} we have implicitly using both the equality $\Box\Box^{-1}=1$, and
$\Box^{-1}\Box=1$. While the former is correct by definition of $\iBox$, the latter is not true since $\Box$ has a non-trivial kernel, given by the functions $f$ such that $\Box f=0$. On such functions the formal equality $\Box^{-1}\Box f=f$ is not true. Thus, to be accurate, we should rather write in \eq{m2s2}
$\sigma\Box^{-1}\Box\sigma$  instead of $\sigma^2$.
Note  that the expression  $\sigma\Box^{-1}\Box\sigma$ has an invariance under constant shifts of the conformal factor, $\sigma\ra\sigma+c$, which is lost if we write it as $\sigma^2$. The equality $\sigma\Box^{-1}\Box\sigma=\sigma^2$ becomes however correct for all functions $\sigma$ such that $\Box\sigma\neq 0$.}
In this sense the term
$R\Box^{-2}R$ is completely analogous to the term (\ref{Fmn2}) in the Yang-Mills case, and its introduction describes a sort of ``condensation" of the conformal mode. The fact that  the term $R\Box^{-2}R$ has this special physical meaning, and the analogy with the YM case, 
makes it more plausible its emergence in the IR limit of $R^2$ gravity.\footnote{However, in $R^2$ gravity the interpretation in terms of the dynamics of the conformal mode is complicated by the fact that the conformal already gets a kinetic term, as well as a mass term,  at tree level from the $R^2$ term, as we discuss in sect.~\ref{sect:conf}. This will eventually lead us to consider, in sect.~\ref{sect:anomaly}, the dynamics of the conformal mode in Einstein gravity supplemented by the anomaly-induced effective action, where no such complication arises, and one can really have a dynamical mass generation for an otherwise massless mode.}

Observe also that the requirement of general covariance does not fix uniquely the term that reduces to $\sigma^2$ to quadratic order in $\sigma$. Indeed, as we already mentioned,  the equations of motion of the 
the models (\ref{RT}) and (\ref{RR}) have the same expansion
 to linear order over Minkowski space, so the corresponding actions are the same in an expansion to  quadratic order over Minkowski, and they both reproduce the term $\propto\sigma^2 $ in the action, although with different non-linear completions. Both models can therefore be in principle justified on the basis of the argument that a mass should be dynamically generated for the conformal mode. 

These arguments therefore suggests that, in the IR
\be
a_2(\Box)\simeq \frac{\Lambda_{\rm RR}^4}{\Box^2}\, ,
\ee
where $\Lambda_{\rm RR}$ is a dynamically generated mass scale,  analogous of the QCD Lambda parameter $\Lambda_{\rm QCD}$.  The parameter $\Lambda_{\rm RR}$
is related to the mass scale $m$ that appears in \eq{RR} by 
\be\label{LRR}
\Lambda_{\rm RR}^4=\frac{1}{12}m^2\mplr^2\, .
\ee
As for the numerical value of $\Lambda_{\rm RR}$,
just as in $\Lambda$CDM the energy fraction $\ola$ is a derived parameter, fixed by the flatness condition, in the non-local model the parameter $m$ is similarly fixed, again by the flatness condition. In $\Lambda$CDM one finds in this way that, to reproduce the observations,  the cosmological constant  must be of order $H_0^2$; not surprisingly, in the non-local model one finds that  $m^2\sim H^2_0$ \cite{Maggiore:2013mea,Maggiore:2014sia,Foffa:2013vma}. Thus, from \eq{LRR}, 
\be
\Lambda^2_{\rm RR}={\cal O}(H_0\mplr)\, . 
\ee
More precisely, using the best-fit values $m\simeq 0.67 H_0$ for model (\ref{RT})~\cite{Maggiore:2013mea}, or $m\simeq 0.28 H_0$ for model (\ref{RR})~\cite{Maggiore:2014sia} we get
$\Lambda_{\rm RR}\simeq 0.4 (H_0\mplr)^{1/2}$
and 
$\Lambda_{\rm RR}\simeq 0.3 (H_0\mplr)^{1/2}$,
respectively.
Thus, numerically, 
\be
\Lambda_{\rm RR}\simeq 0.8\, {\rm meV}\hspace{15mm}  {\rm [model\,\, (\ref{RT})]}\, ,
\ee
and 
\be
\Lambda_{\rm RR}\simeq 0.5\, {\rm meV}\hspace{15mm}  {\rm [model\,\, (\ref{RR})]}\, .
\ee
Finally, we should also in general take into account the effect of the running of the Weyl-squared term in the IR. The argument on the dynamical mass generation for the conformal mode says nothing about the behavior of the form factor for the Weyl-squared term, since the Weyl tensor vanishes for a conformally flat metric. If the form factor $f^2(\Box)$ associated to the 
Weyl-squared term   has no significant IR enhancement, and more generally if it grows slower than $1/\Box^2$ in the far IR,  its contribution to the cosmological dynamics will be negligible  compared to the $R\Box^{-2} R$ term, and we end up exactly with the non-local models of the form (\ref{RT}) or (\ref{RR}) (depending on the precise form of the non-linear terms, since the two models are equivalent at linear order over flat space). In contrast, if also $f^2(\Box)$ develops a singularity proportional to $\Box^{-2}$, a corresponding term proportional to $C_{\mu\nu\rho\sigma}\Box^{-2}C^{\mu\nu\rho\sigma}$ should be included in the cosmological model. This would not affect the background FRW evolution, since $C^{\mu\nu\rho\sigma}$ vanishes in FRW, but would  give a contribution at the level of cosmological perturbations.

\section{A solution to the naturalness problem}\label{sect:sol}

The above mechanism for the emergence of a term $m^2 R\Box^{-2}R$ in the long-distance effective action gives a new perspective on the naturalness problem of the cosmological constant. As is well known, there are two aspects in the cosmological constant problem: the ``old" cosmological constant problem, namely why we do not observe a huge value for the vacuum energy; and the ``new" cosmological constant problem; namely, assuming that some mechanism sets to zero the vacuum energy, what fixes in a technically natural way the observed energy scale associated to dark energy? Here we have nothing to add to the ``old" cosmological constant problem, since we have simply fixed the renormalized value of the cosmological constant to zero.\footnote{A mechanism that could naturally set to zero the value of the vacuum energy density is proposed in \cite{Maggiore:2010wr}.} In contrast, concerning the naturalness problem of the scale associated to dark energy, the above scenario provides a new and interesting viewpoint. In the mechanism that we have proposed,  in $R^2$ gravity the scale $\Lambda_{\rm RR}$ emerges from dimensional transmutation, similarly to $\Lambda_{\rm QCD}$. There is no issue of naturalness for such a quantity, which is determined by the logarithmic running of a dimensionless coupling constant. Of course, the value of the scale generated dynamically in this way cannot be predicted, just as we cannot predict the value of $\Lambda_{\rm QCD}$. The fact that $\Lambda_{\rm RR}$ turns out to be of the order of a  milli-eV is simply a fact of nature, just as $\Lambda_{\rm QCD}$ turns out to be of order 200~MeV.
However, quantities such as  $\Lambda_{\rm QCD}$ and $\Lambda_{\rm RR}$
do not have a power-like dependence on the masses of the particles in the theory, and are therefore technically natural. 

Another attractive aspect of the above construction is that, to explain dark energy,  we do not need to introduce a fundamental particle with an astonishingly small mass $m\sim H_0\sim 10^{-33}$~eV, as is the case for instance in massive gravity. 
In the above construction  the fundamental mass scale of the theory, emerging from dimensional transmutation, is $\Lambda_{\rm RR}$, and in terms of it the effective action  in the IR reads
\be\label{Sg2RinIR}
S_{\rm Mink}\simeq \int d^4x \sqrt{-g}\, \[ \frac{\mplr^2}{2}\,R-\Lambda_{\rm RR}^4R\frac{1}{\Box^2}R\]\, .
\ee
Even if, just as in QCD,  the  value of $\Lambda_{\rm RR}$ cannot be predicted by the model, still the  numerical value required by the comparison with the observation, which turns out to be of the order of the meV, does not look particularly surprising. An extremely small mass scale only emerges when, in \eq{Sg2RinIR}, we factorize  $\mplr^2/2$, rewriting the action as
\be\label{Sg2RinIRbis}
S_{\rm Mink}\simeq  \frac{\mplr^2}{2} \int d^4x \sqrt{-g}\, \[\,R-
 \frac{2\Lambda_{\rm RR}^4}{\mplr^2}R\frac{1}{\Box^2}R\]\, ,
\ee
so the mass scale $m$ defined in \eq{RR} is given by
\be
m=\sqrt{12}\,\, \frac{\Lambda_{\rm RR}^2}{\mplr}\, ,
\ee
which is suppressed with respect to $\Lambda_{\rm RR}$ by an extra factor of order $\Lambda_{\rm RR}/\mplr$, and is then
of order $10^{-33}$~eV. However, the fundamental mass scale generated by the theory is $\Lambda_{\rm RR}$,  not $m$.

\section{Conformal mode dynamics}\label{sect:conf}

It is interesting to understand in more detail the consequences on the conformal mode of a RG flow that runs from 
 $a_1R^2$ at high energies, with $a_1=a_1(\mu)$ a constant, to $\Lambda_{\rm RR}^4 R\Box^{-2}R$ in the IR. 
 We expands on the technical detail in app.~\ref{app:Bard}, and here we summarize the main results. First of all, let us linearize for simplicity over flat space, and restrict to the scalar sector of perturbations. Then the metric can be written in terms of the gauge-invariant Bardeen potentials $\Phi$ and $\Psi$ as 
\be\label{PsiPhitext}
ds^2=-(1-2\Psi) dt^2+(1+2\Phi) d\vx^2\, .
\ee
Equivalently, we can introduce  the conformal factor $\sigma$ and another independent perturbation $\varphi$, defined writing the metric perturbation in the scalar sector in the form
\be\label{sigmavarphi}
ds^2=e^{2\sigma}\[ -(1-2\varphi) dt^2+(1+2\varphi) d\vx^2\]\, .
\ee
At the level of linearized theory, $e^{2\sigma}=1+2\sigma +{\cal O}(\sigma^2)$, so $2\sigma=\Phi-\Psi$ and
$2\varphi=\Phi+\Psi$. In GR, without $R^2$ terms, of course only the tensor modes are dynamical, and scalar perturbations are non-dynamical constrained fields that satisfy a Poisson equation, rather than a Klein-Gordon equation. Indeed, in the presence of matter with energy-momentum tensor $\Tmn$, in GR the Bardeen potentials satisfy
\bees
\n^2\Phi &=&-4\pi G\rho\, ,\label{n2Phirhotext}\\
\n^2\Psi &=&-4\pi G (\rho-2\n^2\Sigma)\, ,\label{n2Psirhotext}
\ees
where $T_{00}=\rho$ is the energy density, and $\Sigma$ the scalar part of the anisotropic stress tensor, see \eqst{T00}{Tij}.
Therefore the conformal mode satisfies
\be\label{n2sigmaSigmatext}
\n^2(\sigma+4\pi G\Sigma)=0\,  .
\ee
Imposing the boundary condition that the $\sigma$ and $\Sigma$ vanish at infinity, the Laplacian is invertible, so we get
$\sigma=-4\pi G\Sigma$. The conformal mode is therefore a constrained field, determined by the matter content, and vanishes if $\Sigma=0$.

The situation changes drastically in $R^2$ gravity. In this case, as we show in app.~\ref{app:Bard}, linearizing over flat space the conformal mode satisfies the equation
\be\label{Boxsigmatext}
(\Box-m_0^2)\(\sigma +4\pi G\Sigma\)=\frac{4\pi G}{3}\, (\rho-3P)\, .
\ee
Thus the conformal mode $\sigma$, shifted by $4\pi G\Sigma$ as in \eq{n2sigmaSigmatext}, now satisfies a massive KG equation, where  $m_0^2$ is just the same as \eq{m02}. Thus, the conformal mode becomes dynamical, and describe the scalar particle that appears in Stelle theory.
As we see from \eq{Boxsigmatext}, its  source is given by the trace of the energy-momentum tensor, $T=\eMN\Tmn=-(\rho-3P)$, as we expect indeed for the conformal mode. 

Consider now the theory with action
\be
S=\int d^4x\sqrt{-g}\, \[ \frac{\mplr^2}{2} R-Ra_1(\Box)R\]\, , 
\ee
where, 
at some UV scale $\mu$, $a_1(\Box)= a_1$ is a constant, while in the IR 
$a_1(\Box)= \Lambda_{\rm RR}^4 \Box^{-2}$. In this case in the UV the conformal mode is described by \eq{Boxsigmatext}. In the IR, in contrast, the model reduces to \eq{RR}. The dynamical content of this theory, in the scalar sector, has been studied
in Sect.~3 of ref.~\cite{Maggiore:2014sia}, where is has been shown that $\Phi$ and $\Psi$ are non-radiative field that satisfy Poisson equations, as in GR. The same is therefore true for the conformal mode $\sigma=(\Phi-\Psi)/2$. We therefore conclude that a RG flows in which $a_1(\Box)$ evolves from a constant in the UV toward $a_1(\Box)= \Lambda_{\rm RR}^4 \Box^{-2}$ in the IR corresponds to a flow in which the conformal mode, which in the UV is dynamical, and is just the spin-0 mode  of $R^2$ theory with mass given by  \eq{m02}, becomes a constrained field as we flow toward the IR.

\section{Dynamical mass generation for the conformal mode in Einstein gravity}\label{sect:anomaly}

The discussion   of the previous sections suggests that, because of IR effects, a mass term could be dynamically generated for the conformal mode.
The introduction of non-local operators of the type $R\Box^{-2}R$ 
allows us just to describe such a  mass term, in a diffeomorphism-invariant manner.

This leads us to investigate more closely the dynamics of the conformal mode in Einstein gravity, without $R^2$ corrections, to see if a similar ``condensation" of the conformal mode might take place. Indeed, there are several indications for the relevance of the conformal mode in GR in the 
IR~\cite{Antoniadis:1991fa,Antoniadis:2006wq}. In classical general relativity the conformal mode is a non-propagating degree of freedom. At the quantum level, when gravity is coupled to massless conformally-invariant fields, it however acquires  a dynamics through the conformal anomaly. As stressed in \cite{Antoniadis:2006wq}, an anomaly implies that the effect of quantum fluctuations can remain large at the longest length and time scales, so quantum fluctuations can affect the behavior of the classical theory even in the far IR.
This is most evident in $D=2$ space-time dimension. In this case, 
the trace anomaly takes the form (see e.g. \cite{Birrell:1982ix}) 
\be\label{traceT}
\langle T^{\mu}_{\mu}\rangle=\frac{N}{24\pi} R\,,
\ee
where $N=N_S+N_F$ is the total number of massless scalar and Dirac fermion fields. The trace anomaly can be derived from the non-local Polyakov action\footnote{The coefficient $N$ in front of  $S_{\rm anom}$  becomes $N-25$ if one also takes into account the metric fluctuations themselves, beside the fluctuations due to matter fields.}
\be
S_{\rm anom}=-\frac{N}{96\pi}\int d^2x\sqrt{-g}\, R\frac{1}{\Box}R\, ,
\ee
whose variation gives the right-hand side of \eq{traceT}, and which can therefore be added to the classical Einstein-Hilbert action, to provide an effective action which takes into account  the quantum effect of the trace anomaly. In $D=2$ one can always write the metric in the form (\ref{gmnsigma}), without truncating the theory, and in terms of the conformal mode the Polyakov action becomes local, 
\be
S_{\rm anom}=\frac{N}{24\pi}\int d^2x\sqrt{-\bar{g}}\, (-\sigma\overline{\square}\sigma+\overline{R}\sigma)\, ,
\ee
where the bars denote quantities computed with the reference metric $\gbmn$. In $D=2$ the effect of the anomaly is especially important, since the Einstein-Hilbert term is a topological invariant and the dynamics is entirely contained in $S_{\rm anom}$. Observe that, in $D=2$,  the propagator of the $\iBox$ operator is logarithmic, $G(x,x')\propto\log (x-x')^2$, and therefore grows with distance.

The situation is conceptually similar in four dimensions.  In this case the trace anomaly is given by (see e.g. \cite{Birrell:1982ix,Buchbinder:1992rb,Shapiro:2008sf} for pedagogical introductions)
\be\label{traceT4d}
\langle T^{\mu}_{\mu}\rangle=b C^2+ b'\(E-\frac{2}{3}\Box R\)+b''\Box R\, ,
\ee
where, as in \eq{SHD}, $C^2$ is the square of the Weyl tensor, $E$ the Gauss-Bonnet term, and the coefficients $b,b',b''$ depends on the number of massless conformally-coupled scalars, massless fermions and massless vector fields. The $\Box R$ term  on the right-hand side of
\eq{traceT4d}
can be derived from the variation of a local action proportional to $\int d^4x\sqrt{-g}\,R^2$. The remaining combination is obtained from the variation of a non-local action
\be
S_{\rm anom}=-\frac{1}{8}\int d^4x\sqrt{-g}\(E-\frac{2}{3}\Box R\) \Delta_4^{-1}
\[ b' \(E-\frac{2}{3}\Box R\)-2b C^2\]\, ,
\ee
where
\be
\Delta_4=\Box^2+2\RMN\n_{\mu}n_{\nu}-\frac{2}{3} R\Box +\frac{1}{3}(\n^{\mu}R)\n_{\mu}
\, .
\ee
Once again, this non-local action becomes local in terms of the conformal mode. Writing the metric as in \eq{gmnsigma},  using as a fiducial metric $\gbmn=\emn$ (and choosing appropriately the local $R^2$ counterterm) one finds 
(see e.g. \cite{Antoniadis:1991fa,Antoniadis:2006wq})
\be\label{Sanom4D}
S_{\rm anom}=-\frac{Q^2}{16\pi^2}\int d^4x\, (\Box\sigma)^2\, ,
\ee
where 
\be\label{Q2} 
Q^2=\frac{1}{180}(N_S+\frac{11}{2}N_F+62 N_V-28)+Q^2_{\rm grav}\, .
\ee
Here $N_S, N_f$ and $N_V$ are the number of massless conformally coupled scalars, massless Weyl fermions and massless vectors, respectively, while $Q^2_{\rm grav}$ is the contribution from the gravitons, which is not unambiguously known, and $-28$ is the contribution from the $\sigma$ field itself~\cite{Antoniadis:1991fa,Antoniadis:1996pb}.

 Of course, a difference from the two-dimensional case is that in $D=4$ writing the metric in the form (\ref{gmnsigma}) is a truncation of the theory. However, we see from \eq{Sanom4D} that the propagator of the $\sigma$ field in momentum space is $\propto 1/k^4$, which in coordinate space, in $D=4$, gives  again a propagator growing logarithmically, as in $D=2$,
\be\label{Gx2m0}
G(x,x')=-\frac{1}{2Q^2}\log\[\mu^2 (x-x')^2\]\, ,
\ee
where $\mu$ is an IR regulator.
This means again that the fluctuations of the $\sigma$ field are large at long distances, and justifies a posteriori the approximation of keeping only the conformal mode when studying the quantum effects of gravity in the IR.
The fact that the conformal mode dominates the quantum theory in the IR is also confirmed by the fact that it gives the most infrared divergent contribution to the propagator in de~Sitter space~\cite{Antoniadis:1986sb}. 
Observe that the fourth-order operator $\Delta_4$  in $D=4$ is the analogue of the second-order operator $\Box$  in $D=2$, since it is conformally invariant, $\sqrt{-g}\,\Delta_4=\sqrt{-\bar{g}}\,\bar{\Delta}_4$.

As discussed in ref.~\cite{Antoniadis:1996pb}, the large fluctuations in the two-point function of the $\sigma$ field due to its logarithmic growth create a situation that looks very similar to what happens in the two-dimensional XY model, with the Berezinsky-Kosterlitz-Thouless (BKT) transition. Let us recall the argument of ref.~\cite{Antoniadis:1996pb}. The four-dimensional theory  for the conformal factor has solutions of the form
\be
\sigma(x)=q\ln |x-x_0|\, ,
\ee
which are analogous to the vortices of the 2-dimensional theory. The euclidean action for $\sigma$, evaluated on this configuration, takes the value~\cite{Antoniadis:1996pb}
\be 
S_E=\frac{1}{2}Q^2q^2\ln\(\frac{L}{a}\)\, ,
\ee
where $L$ is the IR cutoff and the ``lattice spacing" $a$ is the UV cutoff. The parameter $x_0$ is a collective coordinate corresponding to the center of the solution. The number of ways of placing this solution in a volume $V$ is proportional to $V$, so the  corresponding entropy is $4\log L/a$, and the free energy is 
\be
F=\(\frac{1}{2}Q^2q^2-4\)\ln\(\frac{L}{a}\)\, .
\ee
If $Q^2>8/q$ the free energy diverges as we remove the IR cutoff ($L\ra\infty$) or the UV cutoff ($a\ra 0$), so these configurations are irrelevant. In contrast, if $Q^2<8/q$ they dominate the partition function, so at the critical value of $Q$ we expect a phase transition. For the solution with $q=1$,  $Q_c^2=8$. The situation is indeed completely analogous to the BKT phase transition in two dimensions, which is also triggered by the logarithmic growth of the vortex-vortex interaction. The role of the coupling constant to be tuned to get the phase transition is here played by $Q$, which depends on the number of massless conformally-coupled particles. If the unknown graviton contribution in \eq{Q2} is not too large, the theory with just a single massless photon is in the region where $Q^2<Q_c^2$, and the non-trivial configurations dominate.

In ref.~\cite{Antoniadis:1996pb} it has been suggested that  $Q^2<Q_c^2$ corresponds to a phase where the space-time geometry is no longer smooth. Actually, pushing further the analogy with the BKT transition can suggest a different scenario. Indeed, in the two-dimensional case, in the regime where the vortices dominate, the correlation length becomes finite and a mass gap is generated. 
It is plausible to expect that the same happens in the four-dimensional case. 
We therefore expect the generation of a term $\propto \sigma^2$ in the effective action for $\sigma$, whose coefficient is given by a dynamically-generated mass scale. Of course, this mass term must be constructed in such a way that it does not spoil the underlying diffeomorphism invariance, which leads us again to a  term $\Lambda_{\rm RR}^4R\Box^{-2}R$, see \eq{m2s2}.

Adding this term to the  anomaly-induced effective action  (\ref{Sanom4D}), to quadratic order  we get an effective action for $\sigma$ given by
\be\label{Sanom4Dmass}
S^{(2)}_{\rm eff}=-\int d^4x\,\[ \frac{Q^2}{16\pi^2} (\Box\sigma)^2+\Lambda_{\rm RR}^4\sigma^2\]
\, .
\ee
The corresponding euclidean propagator is 
\be\label{Gxmu4}
G(x)=\frac{8\pi^2}{Q^2}\int \frac{d^4k}{(2\pi)^4}\, \frac{1}{k^4+\mu^4} e^{-ikx}\, ,
\ee
where $\mu^4=16\pi^2\Lambda_{\rm RR}^4/Q^2$. Carrying out the angular integrals, we get
\be
G(x)=\frac{2}{Q^2}\int_0^{\infty} dy \frac{y^2 J_1(y)}{y^4+ (\mu^2 x^2)^2}\, ,
\ee
where $J_1(y)$ is a Bessel function.
The final integral can in principle be carried out in term of a (not very illuminating) Meijer's G-function. However we already see from this expression, together with the expansion $J_1(y)\simeq y/2$ for small $y$, that the presence of the term $\mu^2 x^2$ regularizes  the integral near the integration limit $y=0$. In the limit $\mu^2 x^2\ra 0$ we recover \eq{Gx2m0}, except that the artificial IR regulator $\mu$ of \eq{Gx2m0} is  replaced by the quantity $\mu$ that appears in
\eq{Gxmu4}, which now is
a physical quantity related to $\Lambda_{\rm RR}$. In the opposite limit 
$\mu^2 x^2\gg 1$, introducing the notation $u=(\mu^2x^2)^{1/2}$ and using as integration variable $z=y/u$, we have
\bees
G(x)&=&\frac{2}{Q^2}\, \frac{1}{u}\int_0^{\infty}dz\, \frac{z^2}{1+z^4} J_1(uz)\nn\\
&\simeq &\frac{2}{Q^2}\, \sqrt{\frac{2}{\pi}} \, \frac{1}{u^{3/2}}
\int_0^{\infty}dz\, \frac{z^{3/2}}{1+z^4}\, \cos\(uz-\frac{3\pi}{4}\)\, ,
\ees
where in the second line we have used the asymptotic expansion of the Bessel function.\footnote{Observe that, as $u\ra\infty$, the regime $uz$ fixed corresponds to $z\ra 0$, whose contribution to the integral is suppressed by the factor $z^2$ in the integrand. The dominant contribution therefore comes from the region $uz\ra\infty$, where we can use the asymptotic expansion
$J_1(x)\simeq (2/\pi x)^{1/2}\cos (x-3\pi/4)$ valid for large $x$.} The integrand is an oscillating function, which oscillates faster and faster as $u$ increases, so the Green's function vanishes faster than $1/u^{3/2}$, i.e faster than $1/(\mu^2 x^2)^{3/2}$.

The physical picture that emerges from these considerations is that, when the dynamics of the conformal mode is described by the action (\ref{Sanom4D}),  fluctuations grow larger and larger at long distances, as shown by the logarithmic growth of the propagator in \eq{Gx2m0}. This is a sign of an IR instability of the vacuum state. When a mass for the conformal mode is dynamically generated, the long-distance correlations  grow only until the separation $|x-x'|$ reaches a  length-scale  of order $1/\mu$, and after that they decay faster than $1/|x-x'|^3$. Therefore, there is no IR instability. The dynamical generation of a mass for the conformal mode is the mechanism by which the vacuum restructures itself, as a response to the IR instability triggered by the anomaly-induced effective action. 

\section{Conclusions}\label{sect:concl}

In this paper we have explored the possibility of generating non-local terms, such as those given in \eqs{RT}{RR}, through strong IR effects in gravity. We have first of all considered the running of the coupling associated to the $R^2$ term in higher-derivative gravity. We have stressed that a peculiar aspect of the running of this coupling is that, even in the deep IR, its calculation requires a knowledge of the UV completion of GR. Indeed, in GR the corresponding beta function is not even gauge-independent, and has no physical meaning. Stelle theory provides an example of  an extension of GR where, thanks to the improved UV behavior, the beta functions of the ${\cal O}(R^2)$ couplings are physical and gauge-independent. However, the result from Stelle theory cannot be taken literally either, because the theory is plagued by a ghost (and, furthermore, existing computations of the RG flow have been performed only in the regime $E\gg\mplr$). In the absence of a concrete reliable result for the RG flow, we have simply explored the consequences of a scenario in which the coupling $a_2$ associated to the $R^2$ term grows in the IR. In this case, similarly to QCD, we expect the dynamical generation of a mass scale, $\Lambda_{\rm RR}$, and of a non-trivial form factor $a_2(\Box)$. We have provided arguments, based on the analogy with the known non-perturbative behavior  of the gluon propagator in QCD, that suggest that in the strong coupling regime the emergence of a pole-like behavior
$a_2(\Box)\propto\Box^{-2}$ is quite possible. 

The emergence of a dynamical scale $\Lambda_{\rm RR}$ would have far reaching consequences for solving long-standing puzzles concerning dark energy. In particular, this scale emerges from the logarithmic running of a dimensionless coupling constant, similarly to $\Lambda_{\rm QCD}$, and does not receive large loop corrections, contrary to what happens to the vacuum energy. It is therefore technically natural. We found that, to explain dark energy today, 
the dynamically-generated scale $\Lambda_{\rm RR}$ should be of the order of a  milli-eV. Of course we cannot predict its numerical value, just as we 
cannot  predict the value of $\Lambda_{\rm QCD}$. However,  it is interesting that, to obtain a dynamical explanation of dark energy,  we do not need to introduce a fundamental particle with an astonishingly small mass $m\sim H_0\sim 10^{-33}$~eV, nor a cosmological constant $\Lambda$ with the amazingly small value $\Lambda\sim H_0^2$. In our scenario, 
in the IR the action  is given by \eq{Sg2RinIR}, with a fundamental mass scale $\Lambda_{\rm RR}\sim {\rm meV}$, which is by no means a particularly surprising value.
The connection with the present value of the Hubble parameters comes out because, in a cosmological context, derivatives of the metric are of order $H$, so $\Box\sim H^2$ and $\iBox\sim H^{-2}$, while
$R\sim H^2$. Thus 
the term 
$\Lambda^4_{\rm RR}R\Box^{-2}R$ becomes competitive with $\mplr^2R$ when $H\sim \Lambda^2_{\rm RR}/\mplr$ which, for $\Lambda_{\rm RR}\sim\, {\rm meV}$, gives $H\sim H_0\sim 10^{-33}$~eV. However, in this scenario the fundamental scale emerging dynamically in the theory is $\Lambda_{\rm RR}$, and there is no fundamental particle with the astonishingly small mass $10^{-33}$~eV.

Another interesting aspect of the above analysis is that it draws attention to the dynamics of the conformal mode of the metric, since the running of the coupling $a_2$ associated to $R^2$, from a constant value in the UV toward   $a_2(\Box)\propto\Box^{-2}$ in the IR, has an interesting interpretation in terms of the conformal mode, as we discussed in sect.~\ref{sect:conf}. 
This stimulates to look in more detail into the dynamics of the conformal mode in Einstein gravity, in which the higher-derivative term is provided by the anomaly-induced effective action, i.e. by the loops of the massless matter fields in the theory. Once again, the conformal mode seem to play an interesting role. In Einstein gravity supplemented by the anomaly-induced effective action the emergence of a term $R\Box^{-2}R$ can be interpreted as  the dynamical generation a mass term for the conformal mode of the metric (an interpretation that becomes blurred in Stelle theory, where a mass term for the conformal mode is already present at tree level because of the $R^2$ term). We
have argued that such  a dynamical mass generation can take place as a consequence of the strong IR fluctuations generated by the two-point function of the conformal mode, which grows logarithmically at large distances and, as already suggested in 
in ref.~\cite{Antoniadis:1996pb}, can trigger a phase transition similar to  the Berezinsky-Kosterlitz-Thouless transition in two dimensions. If this results in a dynamical mass generation for the conformal mode, one  again obtains in the IR a non-local model of the type (\ref{RT}) or (\ref{RR}).

\vspace{5mm}
\noindent
{\bf Acknowledgments.}
I am grateful to  Eugenio Bianchi, Giulia Cusin,
Eduard Gorbar, Alex Kehagias,  Ilya Shapiro, Sergey Sibiryakov, Julian Sonner and Arkady Tseytlin  for very useful discussions.
This work is supported by the Fonds National Suisse and by the SwissMAP National Center for Competence in Research.

\appendix

\section{Dynamics of $R^2$ gravity in the scalar sector}\label{app:Bard}

In this appendix we give more details on the computation leading to \eq{Boxsigmatext}.
In GR the gauge-invariant scalar and vector perturbations of the metric are non-dynamical, and only tensor perturbations are dynamical. In $R^2$ theory this is different since, as  we mentioned in Sect.~\ref{sect:Stelleth}, there is a scalar particle with a mass $m_0$ given by \eq{m02}. To understand how this takes place, and which scalar perturbation becomes dynamical,
we linearize the metric around flat space, using the  scalar-vector-tensor decomposition of the perturbations. This is the standard tool of  cosmological perturbation theory, but can also be very useful  for separating the perturbations over Minkowski space into pure gauge degrees of freedom, gauge-invariant but non-radiative components, and radiative degrees of freedom. We  first recall how the formalism works in Einstein-Hilbert 
gravity, following~\cite{Flanagan:2005yc,Jaccard:2012ut}, and we will then extend it to $R^2$ gravity.

\subsection{Bardeen variables and conformal factor in GR}

Expanding the metric $\gmn$ over flat space,  $\gmn=\emn+\hmn$,  the metric perturbation can be decomposed as
\bees
h_{00}&=&2\psi\, ,\label{hijpsi}\\
h_{0i}&=&\beta_i+\pa_i\gamma\label{h0ibetagamma}\\
h_{ij}&=&-2\phi\d_{ij}+\(\pa_i\pa_j-\frac{1}{3}\d_{ij}\n^2\)\lambda  +\frac{1}{2}(\pa_i\eps_j+\pa_j\eps_i)
 +h_{ij}^{\rm TT}\, ,\label{hijphilambdaeps}
\ees
where $\beta^i$ and $\eps^i$ are transverse vectors, 
$\pa_i\beta^i=\pa_i\eps^i=0$, and $h_{ij}^{\rm TT}$ is transverse-traceless,
$\pa^jh_{ij}^{\rm TT}=0$ and
$\d^{ij}h_{ij}^{\rm TT}=0$. We have therefore decomposed the 10 components of $\hmn$ into four  functions $\{\psi,\phi,\gamma,\lambda\}$ which are scalar under spatial rotations, two transverse vector $\beta^i$ and $\eps^i$, which carry two independent components each, and a transverse-traceless tensor $h_{ij}^{\rm TT}$, which also has two degrees of freedom.\footnote{In the context of the scalar-vector-tensor decomposition,  by scalar we will always mean scalar under spatial rotations. These quantities  are of course not Lorentz scalars (for instance, $\psi$ is the $(00)$ component of the Lorentz tensor $\hmn$). Similarly for the use of the words vector and tensor.
}
The linearized massless theory is invariant 
under the gauge transformations 
$\hmn\ra\hmn  -(\pam\xin+\pan\xim)$.
It is useful to decompose also the gauge functions $\xi_{\mu}$ in the form
$\xi_0=A$, $\xi_i=B_i+\pa_iC$
where $\pa_i B^i=0$. In terms of these variables the gauge transformation
reads
\bees
&\psi\ra \psi -\dot{A}\, ,\qquad
&\phi\ra \phi+(1/3)\n^2C\, ,\label{tr1}\\
&\gamma\ra\gamma -A-\dot{C}\, ,\qquad
&\lambda\ra\lambda -2C\, ,\label{tr2}\\
&\beta_i\ra\beta_i-\dot{B}_i\, ,\qquad
&\eps_i\ra\eps_i-2B_i\, ,\label{tr3}
\ees
while $h_{ij}^{\rm TT}$ is gauge invariant. One can then check that the combinations
\bees
\Phi&=&-\phi-\frac{1}{6}\n^2\lambda\, ,\label{Bard1}\\
\Psi &=& \psi -\dot{\gamma}+\frac{1}{2}\ddot{\lambda}\, ,\label{Bard2}
\ees
are gauge invariant. Indeed, these are just the Bardeen variables, normally introduced when studying  perturbations over a Friedmann-Robertson-Walker background, restricting to a scale factor $a(t)=1$. Similarly, in the vector sector the transverse vector $\Xi_i=\beta_i-(1/2)\dot{\eps}_i$ is gauge invariant. Thus, out of the 10 components of the metric perturbation $\hmn$, four are pure gauge degrees of freedom, corresponding to the two scalar  functions $A,C$ and the two components of the transverse vector $B_i$. The remaining six degrees of freedom are gauge invariant, and are organized into the two scalars $\Phi$ and $\Psi$, the two components of the transverse vector $\Xi_i$ and the two components of the transverse traceless tensor $h_{ij}^{\rm TT}$. Of course, not all these six physical degrees of freedom are radiative. To show this, we decompose also the energy-momentum tensor as
\bees
T_{00}&=&\rho\, ,\label{T00}\\
T_{0i}&=&S_i+\pa_iS\, ,\label{T0i}\\
T_{ij}&=&P\d_{ij}+\(\pa_i\pa_j-\frac{1}{3}\d_{ij}\n^2\)\Sigma   +\frac{1}{2}(\pa_i\sigma_j+\pa_j\sigma_i)+\sigma_{ij} \, ,
\label{Tij}
\ees
where  $\sigma^i$ and $S^i$ are transverse vectors and
$\sigma_{ij}$ is transverse and traceless. Physically, $\rho$ is the energy density, $P$ the pressure, and $\Sigma$, $\sigma_i$ and $\sigma_{ij}$ are, respectively,  the scalar, vector and tensor part of the anisotropic stress tensor.
These components are not independent, because of energy-momentum conservation. To linear order over flat space energy-momentum conservation reads $\pam\TMN=0$. In particular, in the scalar sector, the equation $\pam T^{\mu 0}=0$ gives
\be\label{n2Srho}
\n^2S=\dot{\rho}\, ,
\ee
while the longitudinal part of $\pam T^{\mu i}=0$  gives~\cite{Jaccard:2012ut}
\be\label{4dotS}
\dot{S}=P+\frac{2}{3}\n^2\Sigma\, .
\ee
At the quadratic level in the action (i.e. at the linearized level in the equations of motion) the scalar, vector and tensor sectors decouple. Let us focus on the scalar sector. Choosing the gauge $\lambda=\gamma=0$, we have 
$\Phi=-\phi$, $\psi=\Psi$, and the metric in the scalar sector can be written as
\be\label{PsiPhi}
ds^2=-(1-2\Psi) dt^2+(1+2\Phi) d\vx^2\, .
\ee
The  quadratic part of the Einstein-Hilbert action, restricted to the scalar sector,  is~\cite{Jaccard:2012ut}
\be\label{SEH2s}
S_{\rm EH}^{(2,s)}=\mplr^2\int d^4x\,  \( -3\dot{\Phi}^2+\pa_i\Phi\pa^i\Phi
-2\pa_i\Phi\pa^i\Psi \)\, .
\ee
To linear order  the interaction term is
\be
S_{\rm int}=\frac{1}{2}\int d^4x\,
  \hmn\TMN\, ,
\ee
which, restricting again to the scalar sector, reads
\be\label{Sints}
S_{\rm int}^{(s)}= \int d^4x \( 3\Phi P+\Psi\rho\)\, .
\ee
We can now find the equations of motion for the scalar variables, by taking the variation of the action
$S^{(s)}=S_{\rm EH}^{(2,s)}+S_{\rm int}^{(s)}$ with respect to $\Phi$ and $\Psi$.\footnote{Observe that, in principle, we should rather take the variation with respect to the original scalar metric component $\phi,\psi,\lambda$ and $\gamma$. However, because of the definitions (\ref{Bard1}) and (\ref{Bard2}), the variation with respect to $\phi$ is the same as that with respect to $\Phi$, and the variation with respect to $\psi$ is the same as that with respect to $\Psi$, while the variations with respect to $\lambda$ and $\gamma$ gives equations that are consequence of the first two, because of gauge invariance, see \cite{Jaccard:2012ut} for details.}
The variation with respect to $\Phi$ gives 
\be\label{24ddotPhi}
\mplr^2\[  6\ddot{\Phi}-2\n^2(\Phi -\Psi)\] =-3 P\, ,
\ee
while the variation with respect to $\Psi$ gives the Poisson equation
\be\label{8n2Phi1}
2\mplr^2\n^2\Phi=- \rho\, .
\ee
\Eq{8n2Phi1} shows that $\Phi$ is a non-dynamical field, which vanishes if $\rho=0$ (since, imposing the boundary condition that $\Phi$ vanishes at infinity, the inversion of the Laplacian is unique, and $\n^2\Phi=0$ implies $\Phi=0$). Equivalently, the initial conditions for $\Phi$ are not free degrees of freedom, but are fixed by \eq{8n2Phi1} in terms of $\rho$, i.e. $\Phi$ is a constrained variables. Since $\Phi$ is uniquely determined by the matter source, the same is true for 
$\ddot{\Phi}$ in \eq{24ddotPhi}. Indeed,
using \eq{8n2Phi1} we also have  $2\mplr^2\n^2\ddot{\Phi}=-\ddot{\rho}$ which, combined with 
\eq{n2Srho} and with the invertibility of the Laplacian, gives $2\mplr^2\ddot{\Phi}=-\dot{S}$. Using also 
\eq{4dotS}, we finally get
\be
\ddot{\Phi}=-\frac{1}{2\mplr^2} \(P+\frac{2}{3}\n^2\Sigma\)\, ,
\ee
which, inserted into \eq{24ddotPhi}, gives a Poisson equation also for $\Psi$. Writing the result in terms of Newton's constant,
with $1/\mplr^2=8\pi G$, we get the standard result
\bees
\n^2\Phi &=&-4\pi G\rho\, ,\label{n2Phirho}\\
\n^2\Psi &=&-4\pi G (\rho-2\n^2\Sigma)\, ,\label{n2Psirho}
\ees
Proceeding similarly in the vector and tensor sectors one finds
\bees
\n^2\Xi_i&=&-16\pi G S_i\, ,\\
\Box h_{ij}^{\rm TT}&=&-16\pi G \sigma_{ij}\label{hijsij}\, .
\ees
This shows that tensor perturbations are indeed radiative degrees of freedom, while the gauge invariant scalar and vector perturbations are non-radiative degrees of freedom that satisfy a Poisson equation, rather than a Klein-Gordon equation. In particular, in the absence of matter, $\Phi=\Psi=0$ and $\Xi_i=0$.

It is also instructive to observe  that, if we incorrectly truncate the theory in the scalar sector setting $\Psi=0$ in the action (\ref{SEH2s}), $\Phi$ would seem a dynamical field, with a fully dynamical equation of motion obtained setting $\Psi=0$ in \eq{24ddotPhi}, and 
furthermore its kinetic term in the action (\ref{SEH2s}) is such that it would seem a  ghost! However, when one correctly keeps both $\Psi$ and $\Phi$, one finds that it is the variation with respect to $\Psi$, \eq{8n2Phi1}, that reduces $\Phi$ to a non-dynamical field.

Finally, let us recast these results in terms of the conformal factor $\sigma$ and another independent perturbation $\varphi$, defined writing the metric perturbation, in the scalar sector, in the form
\be\label{sigmavarphi}
ds^2=e^{2\sigma}\[ -(1-2\varphi) dt^2+(1+2\varphi) d\vx^2\]\, .
\ee
At the level of linearized theory, $e^{2\sigma}=1+2\sigma +{\cal O}(\sigma^2)$. Then, comparing with \eq{PsiPhi}
\bees
2\sigma&=&\Phi-\Psi\, ,\label{2sigma1}\\
2\varphi&=&\Phi+\Psi\, .\label{2sigma2}
\ees
Since $\Phi$ and $\Psi$ are constrained non-dynamical fields, the same is true for $\sigma$ and $\varphi$. In particular, from 
\eqs{n2Phirho}{n2Psirho},
\bees
\n^2\varphi &=&-4\pi G(\rho-\n^2\Sigma)\, ,\label{n2varphi}\\
\n^2\sigma&=&-4\pi G\n^2\Sigma\, .\label{n2sigmaSigma}
\ees
The second equation, together with the invertibility of the Laplacian assured by vanishing boundary conditions at infinity, 
implies $\sigma=-4\pi G\Sigma$. So the conformal factor is proportional to the scalar part of the anisotropic stress tensor, and vanishes in the absence of matter.

\subsection{Bardeen variables and conformal factor in $R^2$ gravity}

We consider now the dynamics in the scalar sector for the action
\be
S_{\rm EH}+S_{R^2}= \int d^4x \sqrt{-g}\, \[ \frac{1}{2}\mplr^2 R -a_2 R^2\]\, .
\ee
Since we are particularly interested in the dynamics of the conformal mode, and the latter drops from the Weyl tensor, which is conformally invariant, we do not consider the $C^2$ term here. Restricting again to the scalar sector and writing again the metric as in \eq{PsiPhi}, to linear order in the perturbations we have 
\be
R=2\(3\ddot{\Phi}-2\n^2\Phi+\n^2\Psi\)\, ,
\ee
Then, the action to quadratic order in the perturbations, in the scalar sector, is given by
\be
S^{(2,s)}=S_{\rm EH}^{(2,s)}+S_{R^2}^{(2,s)}+S_{\rm int}^{(s)}\, , 
\ee
where $S_{\rm EH}^{(2,s)}$ and $S_{\rm int}^{(s)}$
are given in \eqs{SEH2s}{Sints}, respectively, and 
\be 
S_{R^2}^{(2,s)}=-4a_2\int d^4x\, \(3\ddot{\Phi}-2\n^2\Phi+\n^2\Psi\)^2\, .
\ee
We can now repeat the same analysis performed above for GR. 
The variation of the total action with respect to $\Phi$ gives
\be\label{24ddotPhiR2}
\mplr^2\[  6\ddot{\Phi}-2\n^2(\Phi -\Psi)\] -8a_2(3\pa_0^2-2\n^2)(3\ddot{\Phi}-2\n^2\Phi+\n^2\Psi)=-3 P\, ,
\ee
while the variation with respect to $\Psi$ gives
\be\label{8n2Phi1R2}
2\mplr^2\Phi-8a_2(3\ddot{\Phi}-2\n^2\Phi+\n^2\Psi)=- \n^{-2}\rho\, .
\ee
Combining these equations we find 
\be\label{PsipiuPhi}
\mplr^2(\Phi+\Psi)=\Sigma-\n^{-2}\rho\, ,
\ee
and
\be\label{BoxPhiR2}
\(\ddot{\Phi}-\n^2\Phi-\frac{\mplr^2}{12a_2}\Phi\)
=\frac{1}{3\mplr^2}\( \rho-\n^2\Sigma\)+\frac{1}{24a_2}\n^{-2}\rho
\, .
\ee
\Eq{PsipiuPhi} is the same as the equation obtained in GR for the combination $\Phi+\Psi$,  as we see adding \eqs{n2Phirho}{n2Psirho}, while \eq{BoxPhiR2}
is just a Klein-Gordon (KG) equation (with some source term) for a particle with squared mass
$m_0^2=-\mplr^2/(12 a_2)$, in perfect agreement with \eq{m02}. Thus, the field $\Phi$ becomes dynamical. Observe that in the limit $a_2\ra 0$ (at fixed $\mplr$) we only retain the singular terms in \eq{BoxPhiR2}, and we recover the GR result, \eq{n2Phirho}.\footnote{This is completely analogous to what happens in linearized massive gravity, where again $\Phi$ satisfies a massive 
KG equation~\cite{Deser:1966zzb,Alberte:2010it,Jaccard:2012ut}. It is interesting to observe that $\Phi$ and $\Psi$  are scalar only under rotation, but not under boosts, where they  transform in a way dictated, respectively, by  the transformation properties of $h_{00}$ and $h_{ij}$, so they mix with the gauge-invariant variable in the vector sector. Similarly, even  the action in the scalar sector is not  invariant under Lorentz transformations, since boosts mix it with the action in the vector sector, which in turn mixes also  with the tensor sector. The full set of transformations can be found in \cite{Jaccard:2012ut}. Nevertheless once the field $\Phi$ becomes dynamical, its equation of motion nicely collapse into a KG form, as required by the fact that, in the end, its quanta becomes particles that must satisfy the relativistic dispersion relation.}

We can now rewrite the results in terms of the conformal factor $\sigma$ and of the variable $\varphi$, using
\eqs{2sigma1}{2sigma2}. The equation for $\varphi$ is again (\ref{n2varphi}),\footnote{The equation for $\varphi$ will in general get an extra contribution from the $C^2$ term, which is also responsible for generating a dynamical field in the spin-2 sector.} so this remains a constrained variable and satisfies the same equation as in GR, while the equation for $\sigma$ becomes
\be\label{Boxsigma}
(\Box-m_0^2)\(\sigma +4\pi G\Sigma\)=\frac{4\pi G}{3}\, (\rho-3P)\, .
\ee
Thus the conformal mode $\sigma$, shifted by $4\pi G\Sigma$ as in \eq{n2sigmaSigma}, now satisfies a massive KG equation with mass $m_0^2$. Its  source, not unexpectedly, is given by the trace of the energy-momentum tensor, $T=\eMN\Tmn=-(\rho-3P)$.
Thus, the scalar particle that appears in Stelle theory is described by the conformal factor of the metric. For $a_2>0$ we have
$m_0^2<0$, and in this case the conformal factor is  tachyonic.

\bibliographystyle{utphys}
\bibliography{myrefs_massive}

\end{document}